\documentclass[showpacs,aps,prd,nofootinbib,floatfix,amsmath,amssymb]{revtex4}
\usepackage{graphicx}
\usepackage{relsize}
\begin{document}

\makeatletter
%Feynman slash
\newbox\slashbox \setbox\slashbox=\hbox{$/$}
\newbox\Slashbox \setbox\Slashbox=\hbox{\large$/$}
\def\pFMslash#1{\setbox\@tempboxa=\hbox{$#1$}
  \@tempdima=0.5\wd\slashbox \advance\@tempdima 0.5\wd\@tempboxa
  \copy\slashbox \kern-\@tempdima \box\@tempboxa}
\def\pFMSlash#1{\setbox\@tempboxa=\hbox{$#1$}
  \@tempdima=0.5\wd\Slashbox \advance\@tempdima 0.5\wd\@tempboxa
  \copy\Slashbox \kern-\@tempdima \box\@tempboxa}
\def\FMslash{\protect\pFMslash}
\def\FMSlash{\protect\pFMSlash}
\def\miss#1{\ifmmode{/\mkern-11mu #1}\else{${/\mkern-11mu #1}$}\fi}
%%%% Uso:  \pFMSlash{p}
\makeatother

%\tightenlines
\title{Decays $Z\to ggg$ and $Z'\to ggg$ in the minimal $331$ model}
\author{A. Flores-Tlalpa, J. Monta\~no, F. Ram\'\i rez-Zavaleta, and J. J. Toscano}
\address{Facultad de Ciencias F\'{\i}sico Matem\'aticas,
Benem\'erita Universidad Aut\'onoma de Puebla, Apartado Postal
1152, Puebla, Puebla, M\'exico.}
\begin{abstract}
We perform a complete calculation at the one-loop level for the $Zggg$ and $Z'ggg$ couplings in the context of the minimal $331$ model, which predicts the existence of a new $Z'$ gauge boson and new exotic quarks. Bose symmetry is exploited to write a compact and manifest $SU_C(3)$-invariant vertex function for the $Vggg$ ($V=Z,Z'$) coupling. Previous results on the $Z\to ggg$ decay in the standard model are reproduced. It is found that this decay is insensitive to the effects of the new exotic quarks. This in contrast with the $Z'\to ggg$ decay, which is sensitive to both the standard model and exotic quarks, whose branching ratio is larger than that of the $Z\to ggg$ transition by about a factor of $4$.
\end{abstract}


\maketitle

\section{Introduction}
\label{in}There are no couplings of gluons with the neutral electroweak gauge bosons ($V=\gamma, Z$) at the level of classical action in a renormalizable theory\footnote{This class of couplings arises at the level of classical action in the noncommutative standard model~\cite{NCSM}, but this theory is not renormalizable.}, but they can be induced via loops. At the one-loop level, only quartic couplings of the type $Vggg$ and $VVgg$ can be generated, as the trilinear $Vgg$ ones are forbidden at any order of perturbation theory by Yang's theorem~\cite{YT}. In particular, the $Zggg$ coupling is a very interesting prediction of perturbative quantum field theory, which allows one to examine the interplay of the strong interactions and the weak interactions, as it represents a  rare case where purely strongly interacting particles couples to purely weakly interacting particles. Also, this coupling is interesting from the phenomenological point of view because it is much less suppressed than the purely electroweak couplings $VVVV$. Several authors have studied the decay $Z\to ggg$ in the standard model (SM)~\cite{VVVV,AVVVT,AVVVC,TopE}. The Lorentz structure of this vertex is governed by the vector and axial vector couplings of the $Z$ boson to quarks, which leads to an amplitude made of two finite and gauge-invariant subamplitudes that do not interfere among themselves due to their different color structure. Due to this, both the vector and the axial vector subamplitudes characterizing the $Zggg$ coupling have separately been studied in the literature. It turns out to be that, except for some color factors, the vector part of the $Zggg$ is the same as for the four photon interaction in QED~\cite{4Photon}. This result was used in ref.~\cite{Pggg} to calculate the $\gamma^*ggg$ coupling, which further was adapted to study the vector $Zggg$ coupling~\cite{VVVV}. The contribution of triangle diagrams to the axial vector $Zggg$ coupling was calculated in ref.~\cite{AVVVT}, which however is not gauge-invariant. The complete calculation for the axial vector part, which comprise triangle and box diagrams, was done in ref.~\cite{AVVVC}. The impact of the third family is analyzed analytically in the limit $m_b\to 0$ and $m_t\to \infty$ in ref.~\cite{TopE}. In general terms, as we will see below, both the vector and axial vector amplitudes are essentially determined by the third family, the latter one playing a marginal role with respect to the former.

In this work we are interested in studying the rare decays~\cite{PTT} $Z\to ggg$ and $Z'\to ggg$ within the context of the so-called $331$ model~\cite{331}. This model, which is based in the $SU_C(3)\times SU_L(3)\times U_X(1)$ gauge group, predicts the existence of new gauge bosons, among them, a new $Z'$ gauge boson, and has some interesting features~\cite{SOME331PAPERS}, such as the possibility of yielding signals of new physics at the TeV scale. In this model the lepton spectrum is the same as in the SM, but it is arranged in antitriplets of $SU_L(3)$. The quark sector is also arranged in the fundamental representation of this group, which requires the introduction of three new quarks. An interesting feature of the model is that anomalies cancel out when all of the generations are summed over, which means that the family number must be a multiple of the color number, which suggest a possible approach to solving the generation replication problem. In order to endow all the particles with mass, a Higgs sector composed by three triplets and one sextet of $SU_L(3)$ is required, though only one of the triplets is needed to break down $SU_L(3)\times U_X(1)$ into $SU_L(2)\times U_Y(1)$ at the new physics scale $u>v$, with $v\approx 246$ GeV the Fermi scale. In the first stage of spontaneous symmetry breaking (SSB), there emerge singly and doubly charged gauge bosons in a doublet of the $SU_L(2)$ group, as well as a new neutral $Z'$ boson. The three exotic quarks, ($D$ and $S$ with charge $-4/3$ in units of the positron charge, and $T$ with charge $5/3$) do not couple to the $W$ gauge boson since they emerge as singlets of $SU_L(2)$ and get their mass at the $u$ scale. However, these exotic quarks do couple to all the neutral gauge bosons of the theory, namely, $Z'$, $Z$, $\gamma$, and $g$~\cite{T1}. Besides to study the impact of the new quarks on the $Z\to ggg$ decay, we are interested in investigating the peculiarities that could present the $Z'ggg$ couplings due to the presence of these exotic quarks, which are singlets under the $SU_L(2)$ group and present both vector and axial vector couplings to $Z'$. Also, it is interesting to investigate the sensitivity of a new heavy $Z'$ boson to the three standard quark families, as well as to new quark particles. We are motivated by the physics potential of the LHC collider, which will allow one to study directly and in detail the TeV scale region. In particular, the multipurpose ATLAS detector~\cite{ATLAS} has the mission of detecting or exclude the presence of a new $Z'$ boson in the TeV scale. Therefore, it is important to study the decays of this type of particle, including those rare processes, as the $Z'\to ggg$ transition. We will present exact analytical expressions for the corresponding amplitudes, which will be used to reproduce previous results given in the context of the SM for the $Z\to ggg$ decay.

The paper has been organized as follows. In Sec.~\ref{m} a brief description of the minimal $331$ model is presented with emphasis in the neutral currents sector. In Sec.~\ref{cv} the calculation for the one-loop generated on-shell $Vggg$ vertex is presented. Sec.~\ref{rd} is devoted to discuss our results. In Sec.~\ref{co} the results are summarized. Finally, some large mathematical expressions are presented in the Appendix.

\section{The minimal $331$ model}
\label{m} In this section, we will discuss briefly the main features of the $331$ model~\cite{331}, which is based in the $SU_C(3)\times SU_L(3)\times U_X(1)$ gauge group. As already mentioned in the introduction, the lepton sector of the model is the same as in the SM, but it is now arranged as antitriplets of $SU_L(3)$, as follows:
\begin{equation}
L_i=\left( \begin{array}{ll}
l_i \\
\nu_{l_i}\\
l^c_i
\end{array} \right ), \ \ \ (1,3^*,0), \ \ \ i=1,2,3.
\end{equation}
In order to cancel the $SU_L(3)$ anomaly, the same number of fermion triplets and antitriplets are required. This means that two quark families must be accommodate as triplets and the other one as antitriplet. It is customary to choose the third family as the one transforming as antitriplet in order to distinguish the new dynamics effects in the physics of the top quark from that of the lighter families. Accordingly, the three families are specified as follows:
\begin{equation}
Q_{1,2}=\left( \begin{array}{ll}
u \\
d\\
D
\end{array} \right ), \ \  \left( \begin{array}{ll}
c \\
s\\
S
\end{array} \right ), \ \ (3,3,-1/3), \ \ Q_3= \left( \begin{array}{ll}
t \\
b\\
T
\end{array} \right ), \ \ (3,3^*,2/3),
\end{equation}
\begin{equation}
d^c,\ \ s^c, \ \ b^c: \ \ (3^*,1,1/3), \ \ D^c, \ \ S^c: \ \ (3^*,1,4/3),
\end{equation}
\begin{equation}
u^c, \ \ c^c, \ \ t^c: \ \ (3^*,1-2/3), \ \ T^c: \ \ (3^*,1,-5/3),
\end{equation}
where the exotic quarks $D$, $S$, and $T$ have electric charge $-4/3$, $-4/3$, and $5/3$, respectively.

The Higgs sector comprise three triplets and one sextet of $SU_L(3)$, but only one of the triplets is needed to break $SU_L(3)\times U_X(1)$ into $SU_L(2)\times U_Y(1)$. The next stage of spontaneous symmetry breaking (SSB) occurs at the Fermi scale $v$ and is achieved by the remainder two triplets. The sextet is necessary to provide realistic masses for the leptons~\cite{RML}. In the first stage of SSB several particles acquire masses~\cite{SOME331PAPERS,T1}, among them the new $Z'$ gauge boson and the exotic quarks, which are all singlets of $SU_L(2)$ and thus they do not couple to the $W$ gauge boson at the tree level~\footnote{The $\{Z,Z'\}$ basis do not represents indeed mass eingenstates, but it is related to the mass eigenstates $\{Z_1,Z_2\}$ basis through an orthogonal transformation~\cite{T1}. The mixing angle is however very small and can be ignored in the present analysis. }. Many details of the $Z'$ dynamics has already been presented in ref.~\cite{T1}. Very interestingly, in this model the new gauge boson masses are bounded from above~\cite{331,Ng331, T1} due to the theoretical constraint which yields $\sin^2\theta_W=s^2_W\leqslant 1/4$~\cite{331,Ng331}. The fact that the value of $s^2_W$ is very close to $1/4$ at the $m_{Z'}$ scale leads to an upper bound on the scale associated with the first stage of SSB, which translates directly into a bound on the $Z'$ mass given by $m_{Z'}\leqslant 3.1$ TeV~\cite{Ng331}. It turns out to be that when $s^2_W(\mu)=1/4$ the coupling constant $g_X$ associated with the $U_X(1)$ group becomes infinite and a Landau pole arises~\cite{LP}. Here, we will focus on only those features that are relevant for our discussion. In particular, we need the couplings of the $Z$ and $Z'$ gauges bosons to quarks. The neutral currents of  the quark sector of the model can be written as follows~\cite{T1}:
\begin{equation}
{\cal L}^{NC}_{q}=ie\sum_{q}Q_{q}(\bar{q}\gamma_\mu q)A^\mu+\frac{ig}{2c_W}\sum_{q}\Big[\bar{q}\gamma_\mu(g^{q}_{VZ}-g^{q}_{AZ}\gamma_5)qZ^\mu +
\bar{q}\gamma_\mu(g^{q}_{VZ'}-g^{q}_{AZ'}\gamma_5)qZ'^\mu \Big],
\end{equation}
where the electromagnetic current has been included too. The intensity of the diverse couplings are presented in Table \ref{TABLE}. In this table, $s_W(c_W)$ stands for $\sin\theta_W(\cos\theta_W)$ of the weak angle. On the other hand, the Feynman rules of QCD are well-known, so we are ready to calculate the amplitude for the on-shell $Vggg$ ($V=Z,Z'$) vertex. This will be carried out in the next section. It should be mentioned that there is a different version of this model~\cite{PT} which introduces exotic leptons but with the same quark sector. Since in both versions the model the quark sector is accommodate in the same representation of the $SU_L(3)\times U_X(1)$ gauge group, our results are also applicable to this version with exotic leptons.

\begin{table}
 \caption{\label{TABLE} Structure of the neutral currents for the quark sector of the minimal $331$ model.}
 \begin{ruledtabular}
 \begin{tabular}{|c|c|c|c|c|c|}
  Quark & $Q_{q}$ & $g_{VZ}^{q}$ & $g_{AZ}^{q}$ & $g_{VZ'}^{q}$ & $g_{AZ'}^{q}$ \\
   &  &  &  &  & \\
  \hline
  &  &  &  &  & \\
  $u,c$ & $+\frac{2}{3}$ & $\frac{3-8 s_W^2}{6}$ & $\frac{1}{2}$ & $-\frac{1-6s_W^2}{2\sqrt{3}\;c_W^2\sqrt{1-4s_W^2}}$ & $-\frac{1+2s_W^2}{2\sqrt{3}\;c_W^2\sqrt{1-4s_W^2}}$ \\
 \hline
 &  &  &  &  & \\
  $d,s$ & $-\frac{1}{3}$ & $-\frac{3-4s_W^2}{6}$ & $-\frac{1}{2}$ & $-\frac{1}{2\sqrt{3}\;c_W^2\sqrt{1-4s_W^2}}$ &
  $-\frac{\sqrt{1-4s_W^2}}{2\sqrt{3}\;c_W^2}$ \\
  \hline
  &  &  &  &  & \\
  $D,S$ & $-\frac{4}{3}$ & $\frac{8 s_W^2}{3}$ & $0$ & $\frac{1-9s_W^2}{\sqrt{3}\;c_W^2\sqrt{1-4s_W^2}}$ & $\frac{1}{\sqrt{3}\sqrt{1-4s_W^2}}$ \\
  \hline
  &  &  &  &  & \\
  $b$ & $-\frac{1}{3}$ & $-\frac{3-4s_W^2}{6}$ & $-\frac{1}{2}$ & $\frac{1-2s_W^2}{2\sqrt{3}\;c_W^2\sqrt{1-4s_W^2}}$ & $\frac{1+2s_W^2}{2\sqrt{3}\;c_W^2\sqrt{1-4s_W^2}}$ \\
  \hline
  &  &  &  &  & \\
  $t$ & $+\frac{2}{3}$ & $\frac{3-8s_W^2}{6}$ & $\frac{1}{2}$ & $\frac{1+4s_W^2}{2\sqrt{3}\;c_W^2\sqrt{1-4s_W^2}}$ & $\frac{\sqrt{1-4s_W^2}}{2\sqrt{3}\;c_W^2}$ \\
  \hline
  &  &  &  &  & \\
  $T$ & $+\frac{5}{3}$ & $-\frac{10 s_W^2}{3}$ & $0$ & $-\frac{1-11s_W^2}{\sqrt{3}\;c_W^2\sqrt{1-4s_W^2}}$ & $-\frac{1}{\sqrt{3}\sqrt{1-4s_W^2}}$
\end{tabular}
\end{ruledtabular}
\end{table}

\section{The one-loop $Vggg$ coupling}
\label{cv}In this section, we present the calculation for the on-shell $Vggg$ ($V=Z,Z'$) vertex. Since the Lorentz structure of the neutral currents is the same for both the $Z$ and $Z'$ gauge bosons, we will present a generic amplitude for the $Vggg$ vertex. We will present explicit expressions for this amplitude in terms of Passarino-Veltman scalar functions~\cite{PV}. To begin with, we establish our notation and conventions. The momenta, Lorentz indices, and color indices are defined as follows:
\begin{equation}
V_{\mu_4}(p_4)g^a_{\mu_1}(p_1)g^b_{\mu_2}(p_2)g^c_{\mu_3}(p_3),
\end{equation}
where all momenta are taken incoming. We will present our results in terms of scalar products of the way $p_i\cdot p_j\equiv p_{ij}$, which are adequate to discuss both of the related processes, namely, the $V\to ggg$ decay, which is the purpose of this work, and the $gg\to gV$ reaction, which will be reported in a future communication together with the processes $gg\to \gamma Z$, $gg\to \gamma Z'$, and $gg\to ZZ'$~\cite{WIP}.

\begin{figure}
\centering
\includegraphics[width=2.0in]{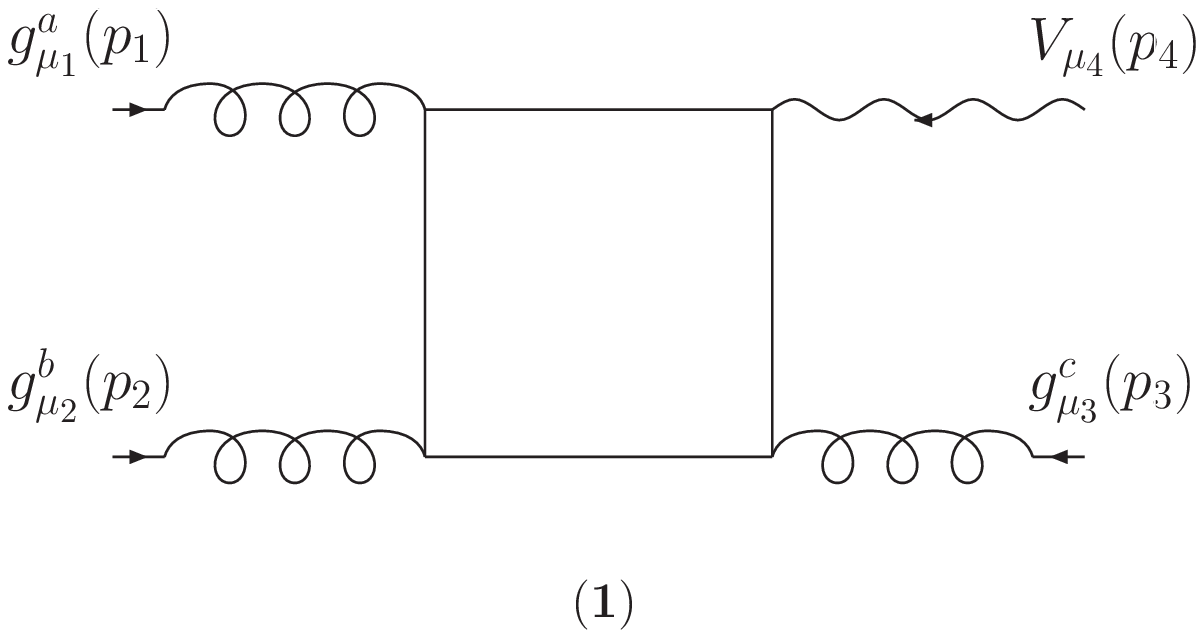}
\includegraphics[width=2.0in]{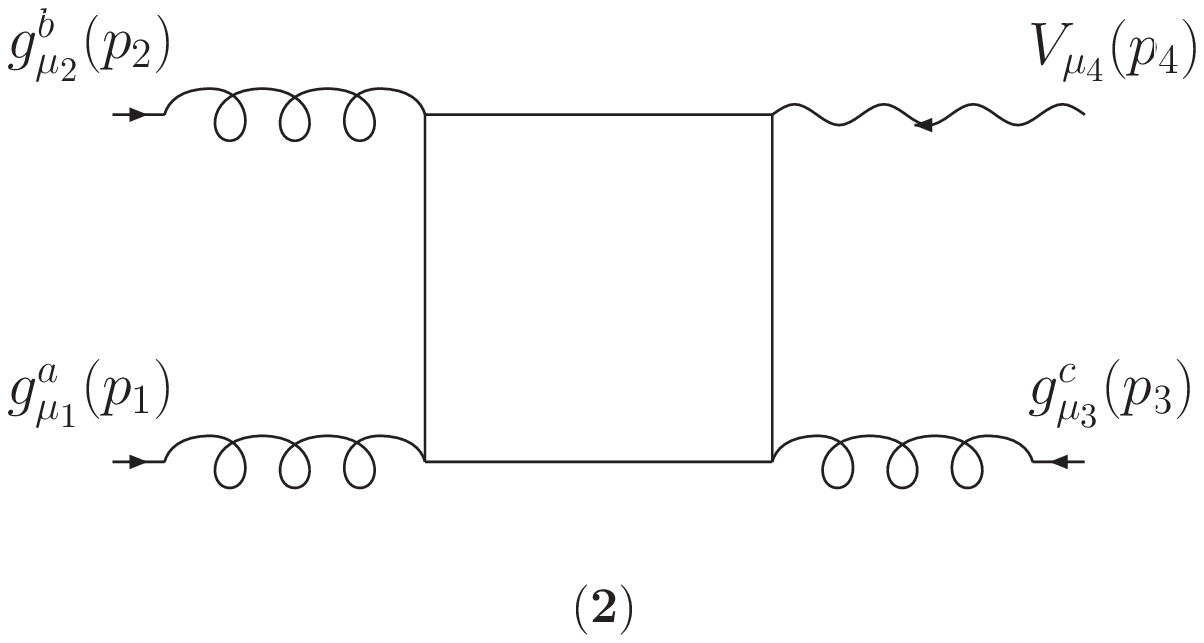}
\includegraphics[width=2.0in]{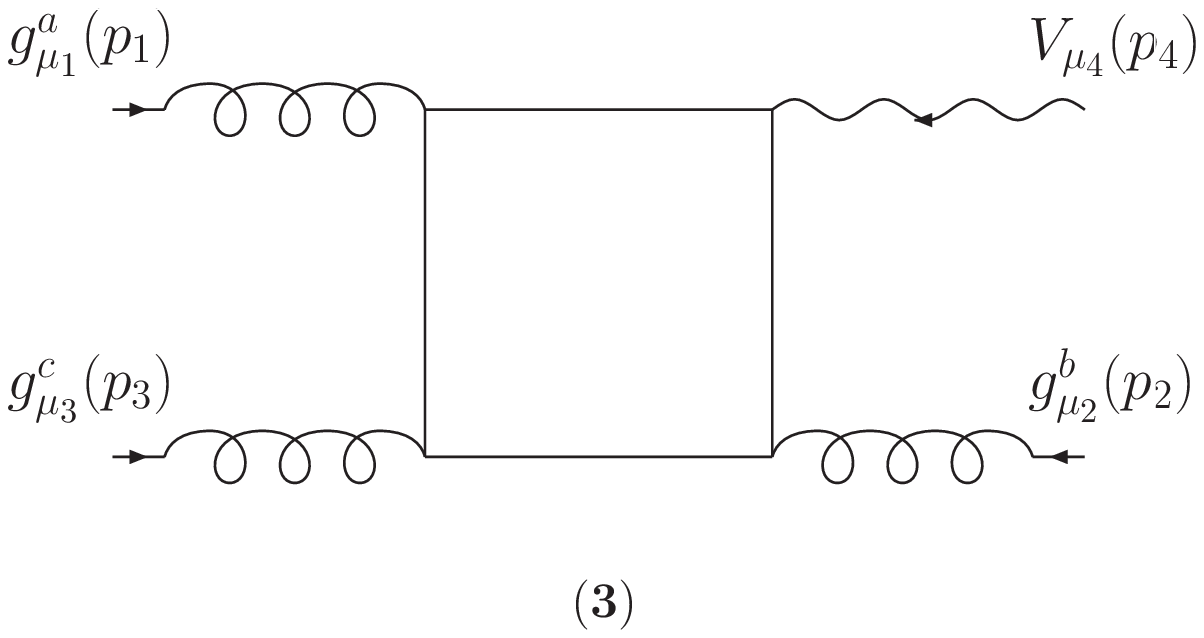}
\includegraphics[width=2.0in]{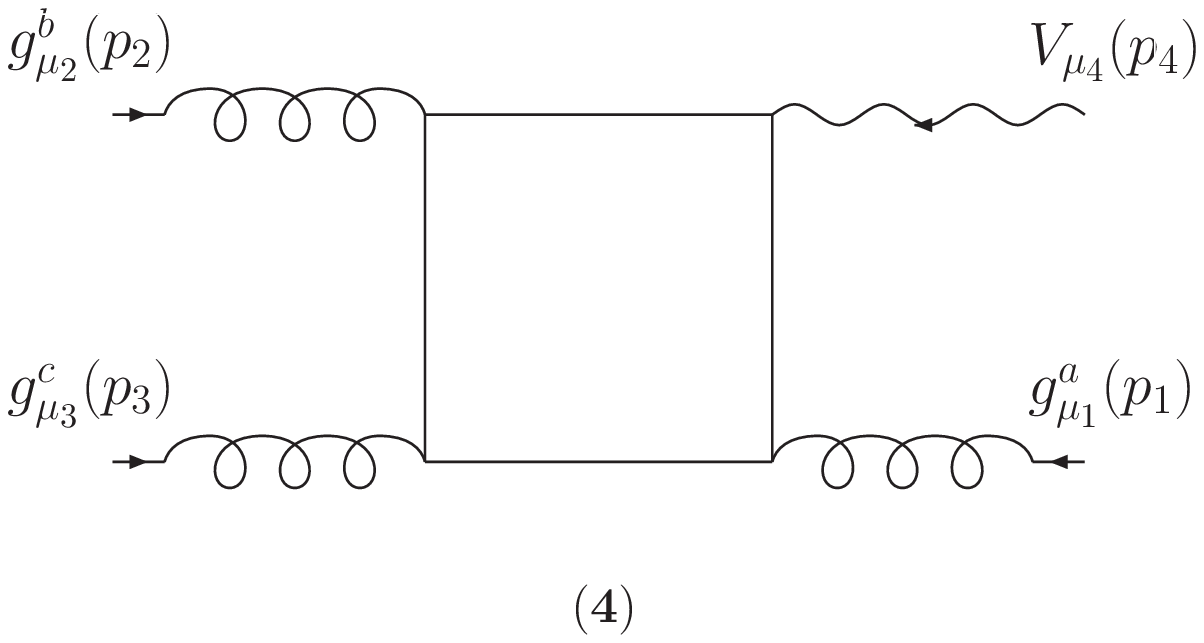}
\includegraphics[width=2.0in]{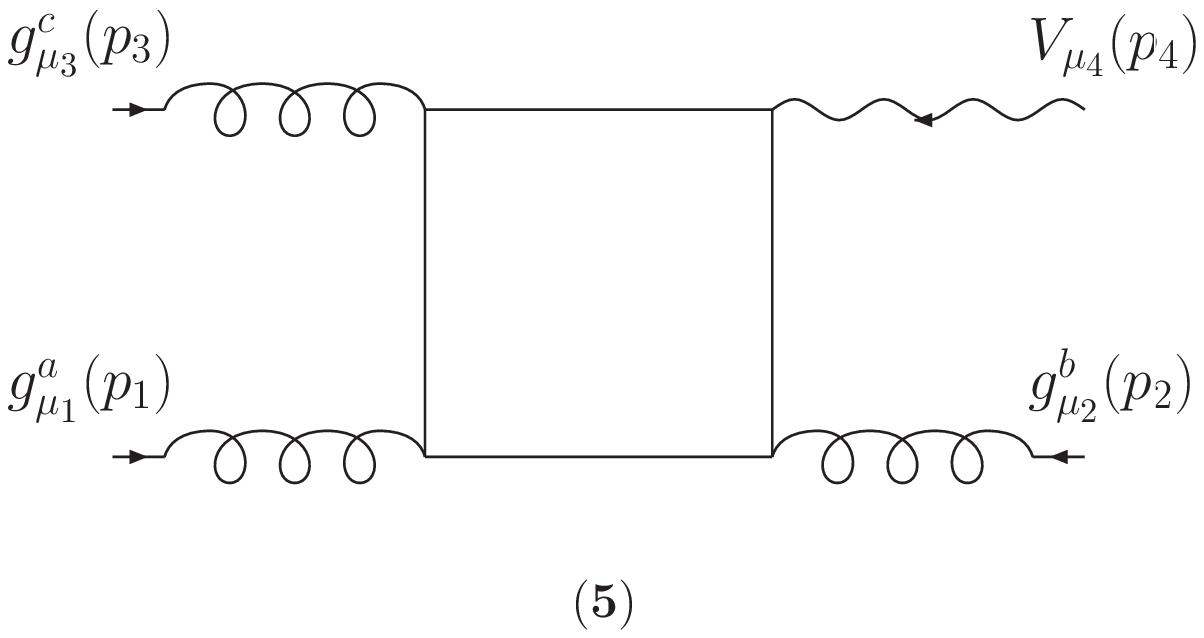}
\includegraphics[width=2.0in]{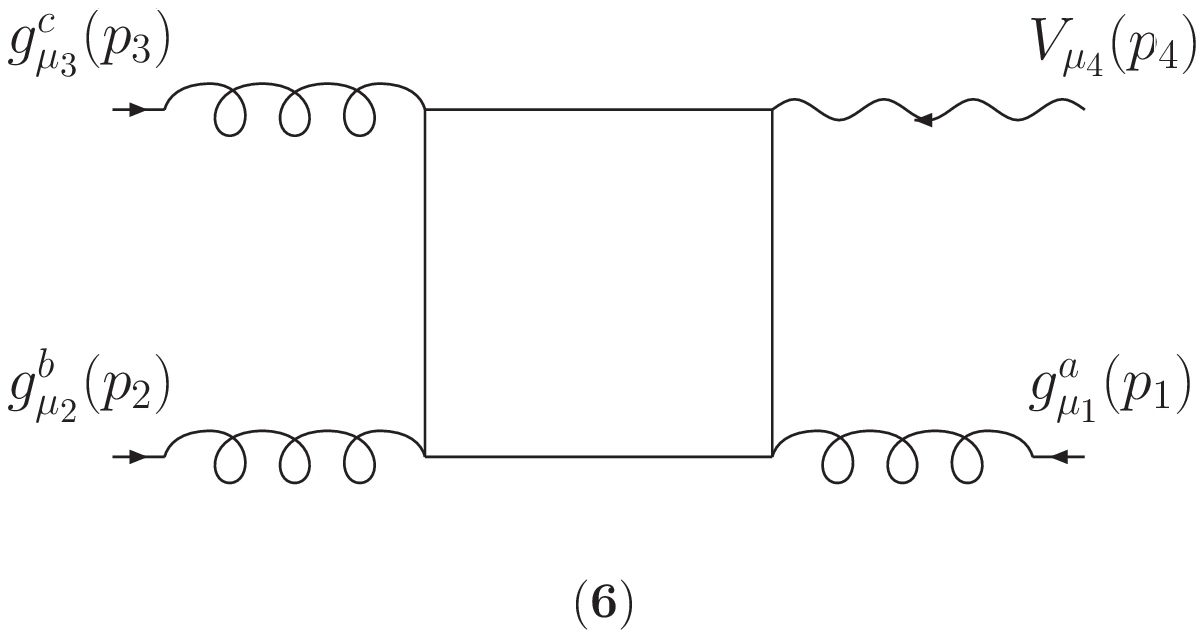}
\caption{\label{Boxes}Box diagrams contributing to the $Vggg$ vertex.}
\end{figure}

\begin{figure}
\centering
\includegraphics[width=2.0in]{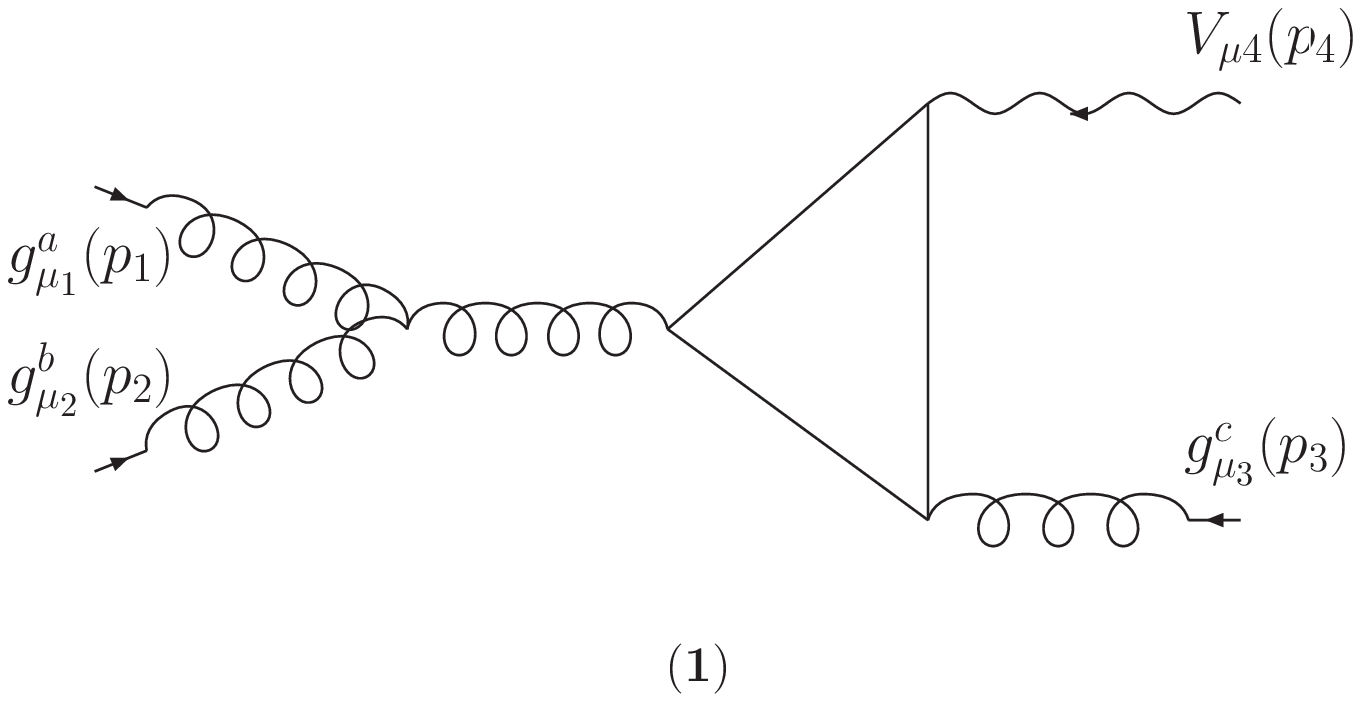}
\includegraphics[width=2.0in]{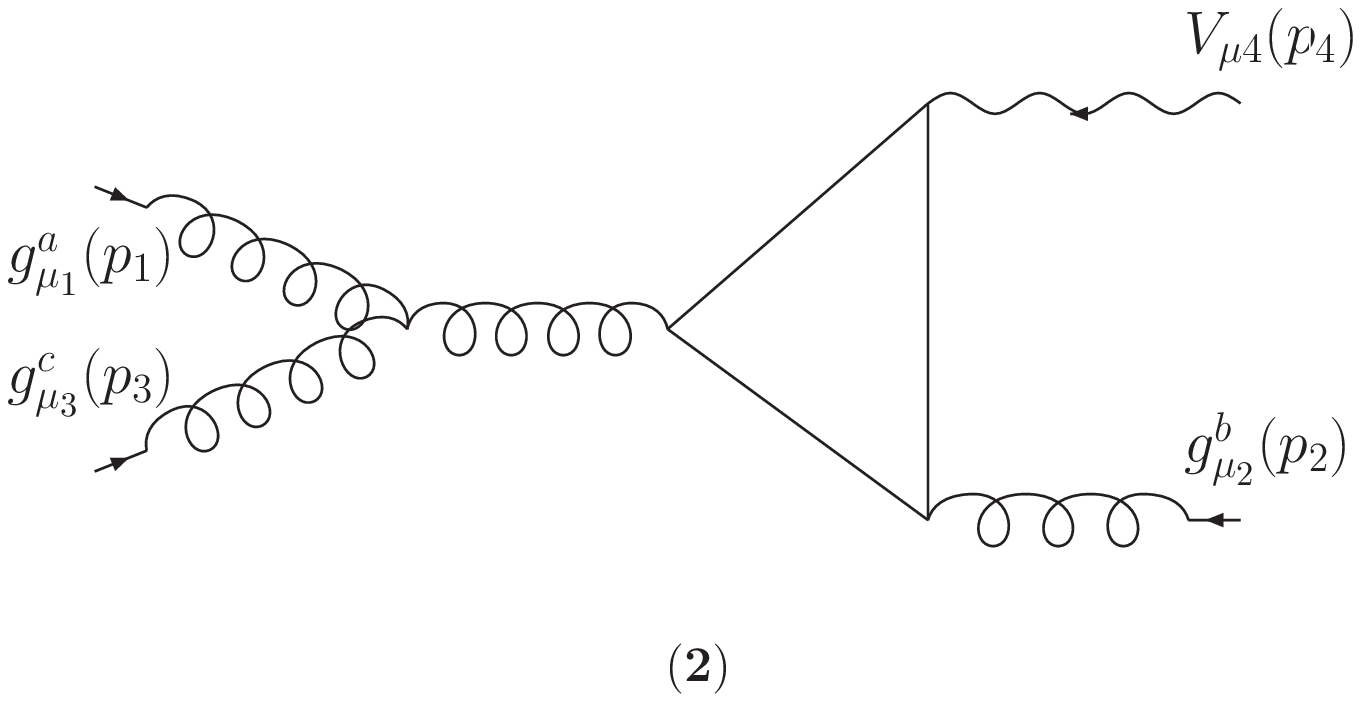}
\includegraphics[width=2.0in]{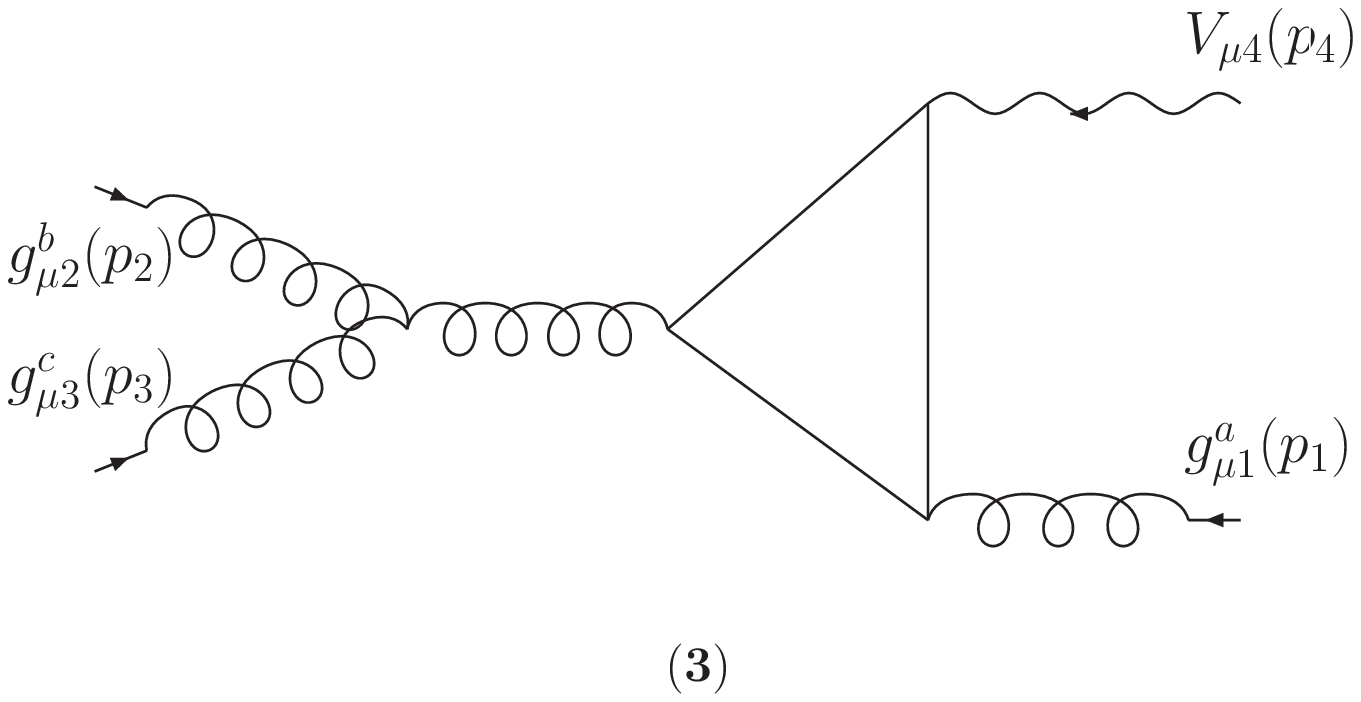}
\includegraphics[width=2.0in]{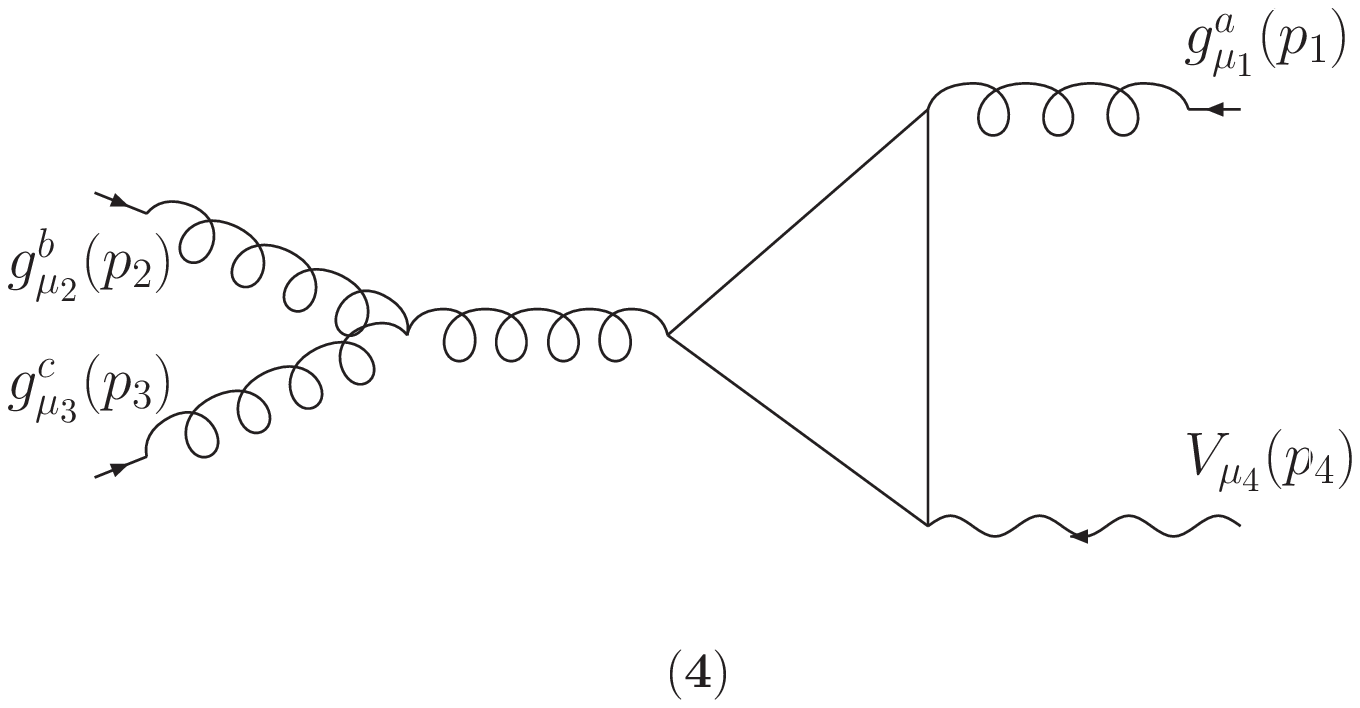}
\includegraphics[width=2.0in]{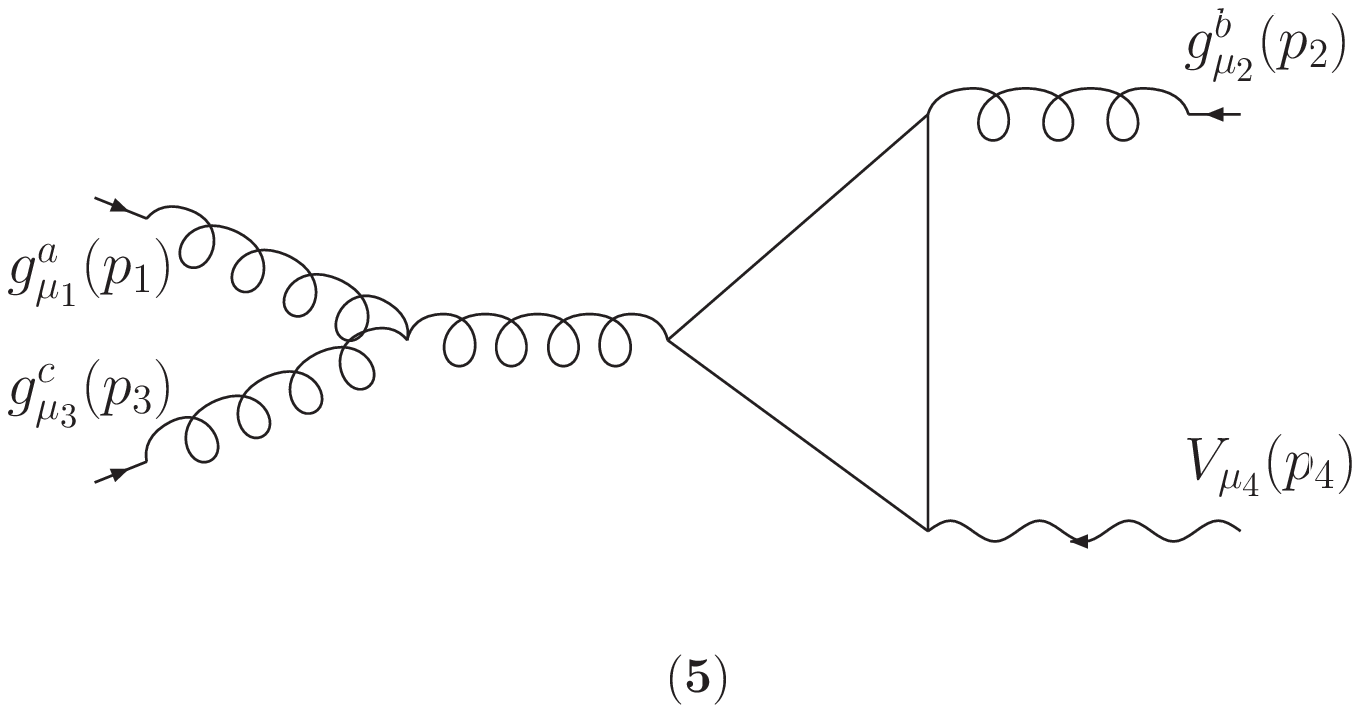}
\includegraphics[width=2.0in]{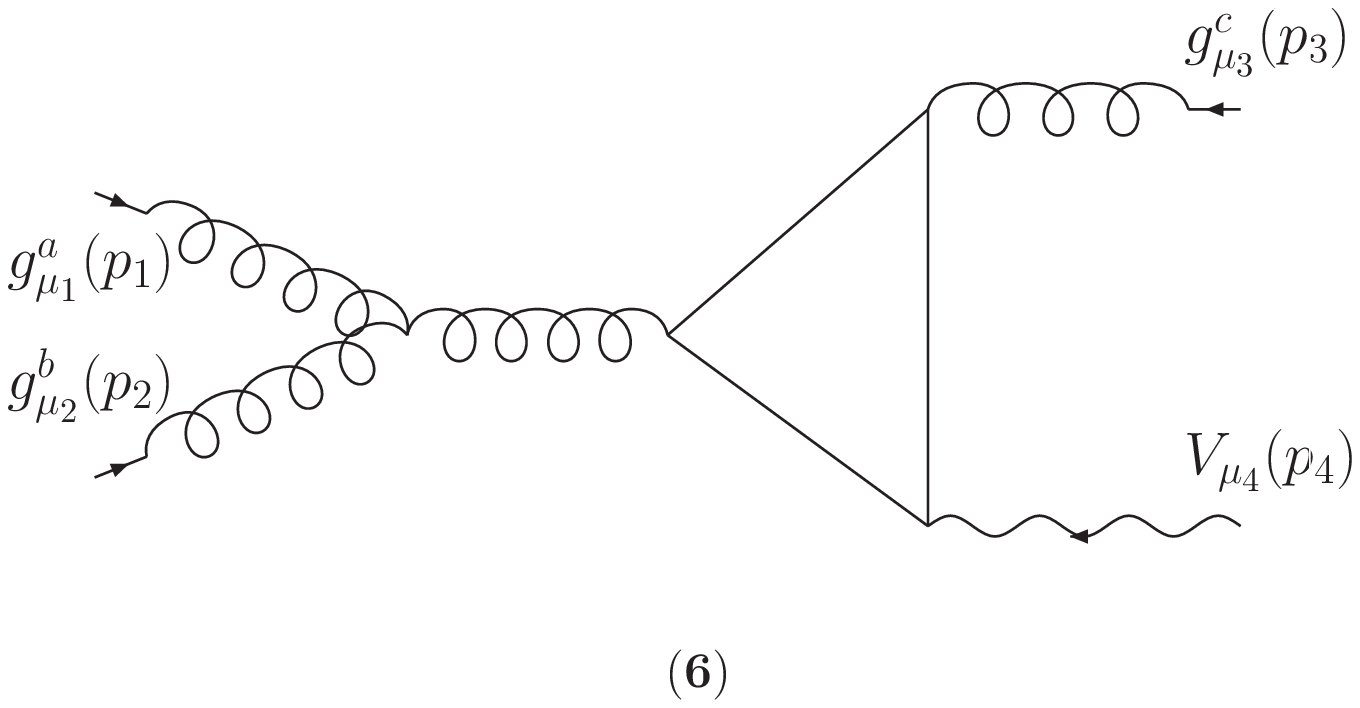}
\caption{\label{Triangles}Triangle diagrams contributing to the $Vggg$ vertex.}
\end{figure}

We now proceed to describe the calculation. The contribution to the $Vggg$ coupling occurs through box and triangle diagrams, which are shown in Fig.~\ref{Boxes} and Fig.~\ref{Triangles}, respectively. There are six box diagrams and six triangle diagrams, but only is needed work out one of each class, as the rest are related by Bose symmetry. The invariant amplitude can be written as follows:
\begin{equation}
{\cal M}_{Vggg}=\sum_{q}{\cal M}^{\mu_1 \mu_2 \mu_3 \mu_4}_{abc}\epsilon^a_{\mu_1}(p_1,\lambda_1)\epsilon^b_{\mu_2}(p_2,\lambda_2)\epsilon^c_{\mu_3}(p_3,\lambda_3)\epsilon_{\mu_4}(p_4,\lambda_4)\; ,
\end{equation}
where the sum is over all quark flavors. This amplitude in turns can be separated into two components as follows
\begin{equation}
{\cal M}^{\mu_1 \mu_2 \mu_3 \mu_4}_{abc}={\cal M}^{\mu_1 \mu_2 \mu_3 \mu_4}_{B\;abc}+{\cal M}^{\mu_1 \mu_2 \mu_3 \mu_4}_{T\;abc},
\end{equation}
with $B$ and $T$ stand for box and triangle contributions. The Lorentz tensor structure of the amplitude is dictated by color gauge invariance and Bose symmetry. Gauge invariance means that the amplitude must satisfies the following transversality conditions
\begin{equation}
p_{i\mu_i}{\cal M}^{\mu_1 \mu_2 \mu_3 \mu_4}_{abc}=0,  \ \ \ i=1,2,3,
\end{equation}
whereas Bose symmetry requires that ${\cal M}^{\mu_1 \mu_2 \mu_3 \mu_4}_{abc}$ must be symmetric under the interchange of both $i\leftrightarrow j$ $(i,j=1,2,3)$ and color indexes. The contribution from the box diagrams displayed in Fig.~\ref{Boxes} can be written as
\begin{eqnarray}
\label{ib}
\mathcal{M}_{B\;abc}^{\mu_1\mu_2\mu_3\mu_4} &=&
\sum_{i=1}^6 \mathcal{F}_i\;\mathcal{I}_{B\; i}^{\mu_1\mu_2\mu_3\mu_4} ,
\end{eqnarray}
where
\begin{eqnarray}
\mathcal{F}_{1,4,5}&\equiv&-g_s^3 g_VN_C\frac{1}{4}(d_{abc}+i f_{abc})\; , \\
\mathcal{F}_{2,3,6}&\equiv&-g_s^3 g_VN_C\frac{1}{4}(d_{abc}-i f_{abc})\; ,
\end{eqnarray}
where $d_{abc}$ and $f_{abc}$ are the totally symmetric and totally antisymmetric structure constants of the color group. The color structure constants can be obtained from the commutation relations $[T^a,T^b]=i\,f_{abc}T^c$ and the anticommutation relations $\{T^a,T^b\}=\delta^{ab}/3+d_{abc}T^c$ for the $SU_C(3)$ generators. In addition, $g_V=g/2c_W$ and $N_C=3$ is the quark color number. The $\mathcal{I}_{B\; i}^{\mu_1\mu_2\mu_3\mu_4}$ tensors appearing in the above expression are given by
\begin{equation}\label{}
\mathcal{I}_{B\;i}^{\mu_1\mu_2\mu_3\mu_4}=
 \int\frac{d^Dk}{(2\pi)^D} \frac{T^{\mu_1\mu_2\mu_3\mu_4}_{B\;i}}
{\Delta_{B\;i}} \;,
\end{equation}
where
\begin{eqnarray}
  T_{B\;1}^{\mu_1\mu_2\mu_3\mu_4} &=&
  \mathrm{Tr} \left\{\gamma^{\mu_4}(g_{VV}^{q}-g_{AV}^{q}\gamma^5)(\pFMSlash{k}+m_{q})\gamma^{\mu_1}
  [(\pFMSlash{k}-\pFMSlash{p}_1)+m_{q}]\gamma^{\mu_2}[(\pFMSlash{k}-\pFMSlash{p}_1-\pFMSlash{p}_2)+m_{q}]
  \gamma^{\mu_3} \right. \nonumber \\
  &&  \left. \times [(\pFMSlash{k}-\pFMSlash{p}_1-\pFMSlash{p}_2-\pFMSlash{p}_3)+m_{q}] \right\} \;,
  \end{eqnarray}
  \begin{eqnarray}
  \Delta_{B\;1} &=& (k^2-m_{q}^2)[(k-p_1)^2-m_{q}^2][(k-p_1-p_2)^2-m_{q}^2][(k-p_1-p_2-p_3)^2-m_{q}^2]
   \;.
\end{eqnarray}
The remainder 5 box integrals can be obtained by Bose symmetry as illustrated in Fig.~\ref{Boxes}.

On the other hand, the contribution arising from the triangle diagrams given in Fig.~\ref{Triangles} can be written as follows:
\begin{equation}
\mathcal{M}_{T\; abc}^{\mu_1\mu_2\mu_3\mu_4} =
\sum_{i=1}^6 \mathcal{F}'_i\;\mathcal{I}_{T\;i}^{\mu_1\mu_2\mu_3\mu_4} \;,
\end{equation}
where
\begin{eqnarray}
\mathcal{F}'_{1,3,4,6}&=&-g_s^3 g_V N_C\left(-\frac{i}{2} f_{abc}\right) \; ,\\
  \mathcal{F}'_{2,5}&=&-g_s^3 g_V N_C\left(\frac{i}{2} f_{abc}\right) \;.
\end{eqnarray}
In the above expression,
\begin{equation}\label{}
\mathcal{I}_{T\;i}^{\mu_1\mu_2\mu_3\mu_4}=
\int\frac{d^Dk}{(2\pi)^D} \frac{T^{\mu_1\mu_2\mu_3\mu_4}_{T\;i}}
{\Delta_{T\;i}} \;,
\end{equation}
where
\begin{eqnarray}
T_{T\;1}^{\mu_1\mu_2\mu_3\mu_4} &=&
  \mathrm{Tr} \left\{\gamma^{\mu_4}(g_{VV}^{q}-g_{AV}^{q}\gamma^5)(\pFMSlash{k}+m_{q})\gamma^{\omega}
  [(\pFMSlash{k}-\pFMSlash{p}_1-\pFMSlash{p}_2)+m_{q}]\gamma^{\mu_3} [(\pFMSlash{k}-\pFMSlash{p}_1-\pFMSlash{p}_2-\pFMSlash{p}_3)+m_{q}] \right\}\nonumber \\
  &&\times \frac{1}{(p_1+p_2)^2}\left[ g_{\omega\rho}+(\xi-1)\frac{(p_1+p_2)_\omega(p_1+p_2)_\rho}{(p_1+p_2)^2} \right]
  [g^{\mu_2\mu_1}(p_2-p_1)^\rho+g^{\mu_1\rho}(2p_1+p_2)^{\mu_2}\nonumber \\
  &&-g^{\rho\mu_2}(p_1+2p_2)^{\mu_1}]\;,
  \end{eqnarray}
  \begin{eqnarray}
  \Delta_{T\;1} &=&(k^2-m_{q}^2)[(k-p_1-p_2)^2-m_{q}^2][(k-p_1-p_2-p_3)^2-m_{q}^2]
   \;.
\end{eqnarray}
As in the box diagrams case, the remainder 5 triangle integrals can be obtained by Bose symmetry as illustrated in Fig.~\ref{Triangles}.

Notice that we have introduced the general propagator for the virtual gluon, which depends on the gauge parameter $\xi$. However, the amplitude is gauge-independent, as the longitudinal component of the gluon propagator does not contribute. To solve the above integrals, we have used the Passarino-Veltman tensorial decomposition \cite{PV} implemented in the FeynCalc computer program \cite{FC}.

Once solved the loop integrals, the amplitude can be expressed as the sum of the vector part and the axial vector part as follows
\begin{equation}
\mathcal{M}_{abc}^{\mu_1\mu_2\mu_3\mu_4} =
\mathcal{M}^{\mu_1\mu_2\mu_3\mu_4}_{V\;abc}+
\mathcal{M}^{\mu_1\mu_2\mu_3\mu_4}_{A\;abc} \;.
\end{equation}
The vector amplitude $\mathcal{M}^{\mu_1\mu_2\mu_3\mu_4}_{V\;abc}$ receives contributions only from box diagrams, whereas the axial vector amplitude $\mathcal{M}^{\mu_1\mu_2\mu_3\mu_4}_{A\;abc}$ receives contributions from both box diagrams and triangle diagrams. Both amplitudes satisfy separately the transversality conditions:
\begin{eqnarray}
p_{i\mu_i}\mathcal{M}^{\mu_1\mu_2\mu_3\mu_4}_{V\;abc}&=&0 \; ,\; i=1,2,3,4 \; ,\\
p_{i\mu_i}\mathcal{M}^{\mu_1\mu_2\mu_3\mu_4}_{A\;abc}&=&0 \; ,\;  i=1,2,3\; .
\end{eqnarray}
Notice that the vector amplitude also satisfies transversality conditions for the $V$ vector boson. It is important to comment that the axial vector amplitude is transverse only after summing over the box and triangle diagrams contributions. Also, each type of diagrams leads to a finite amplitude, \textit{i.e.}, the contributions from box and triangle diagrams to the axial vector amplitude are separately finite. Also, Bose symmetry is satisfied separately by each type of diagrams:
\begin{eqnarray}
\mathcal{M}^{\mu_1\mu_2\mu_3\mu_4}_{V,AB,AT\;abc} &=&
\mathcal{M}^{\mu_1\mu_2\mu_3\mu_4}_{V,AB,AT\;abc}(p_1,\mu_1,a\leftrightarrow p_2,\mu_2,b) \nonumber \\
&=&
\mathcal{M}^{\mu_1\mu_2\mu_3\mu_4}_{V,AB,AT\;abc}(p_1,\mu_1,a\leftrightarrow p_3,\mu_3,c) \\
&=&\mathcal{M}^{\mu_1\mu_2\mu_3\mu_4}_{V,AB,AT\;abc}(p_2,\mu_2,b\leftrightarrow p_3,\mu_3,c)  \nonumber \;.
\end{eqnarray}
where $V,AB,AT$ stand for vector contribution, axial contribution from box diagrams, and axial contribution from triangle diagrams. On the other hand, while the vector amplitude is proportional to $d_{abc}$, the axial amplitude is proportional to $f_{abc}$. Accordingly, the vector amplitude can be written as
\begin{equation}
\mathcal{M}_{V\;abc}^{\mu_1\mu_2\mu_3\mu_4} = g_{VV}^{q}d_{abc}\left(-\frac{ig_s^3g_VN_C}{4\pi^2}\right)\sum_{j=1}^{18} f^{q}_{V_j} T_{V_j}^{\mu_1\mu_2\mu_3\mu_4} \;,
\end{equation}
where the $f^{q}_{V_j}$ are finite form factors given in terms of Passarino-Veltman scalar functions, which are listed in the Appendix. The $T_{V_j}^{\mu_1\mu_2\mu_3\mu_4}$ Lorentz tensors are gauge structures, \textit{i.e.}, they satisfy
\begin{equation}
    p_{j\mu_j} T_{V_j}^{\mu_1\mu_2\mu_3\mu_4} =0 \; , \; j=1,2,3,4 \;.
\end{equation}
The set of 18 terms $f^{q}_{V_j} T_{V_j}^{\mu_1\mu_2\mu_3\mu_4}$ appearing in the vector amplitude, can be divided into 3 subsets, each composed of 6 members,  all them related amongst themselves by Bose symmetry. These subsets can conveniently be organized as follows:
\begin{eqnarray}
&&\{f^{q}_{V_1} T_{V_1}^{\mu_1\mu_2\mu_3\mu_4}, \cdots , f^{q}_{V_6} T_{V_6}^{\mu_1\mu_2\mu_3\mu_4}\}\; , \nonumber \\
&&\{f^{q}_{V_7} T_{V_7}^{\mu_1\mu_2\mu_3\mu_4}, \cdots , f^{q}_{V_{12}} T_{V_{12}}^{\mu_1\mu_2\mu_3\mu_4}\}\; , \nonumber \\
&&\{f^{q}_{V_{13}} T_{V_{13}}^{\mu_1\mu_2\mu_3\mu_4}, \cdots , f^{q}_{V_{18}} T_{V_{18}}^{\mu_1\mu_2\mu_3\mu_4}\}\;. \nonumber
\end{eqnarray}
In this way, it is only necessary to list one element of each set, for instance, the first one of each subset. Making this choice, the respective gauge structures can be written as
\begin{eqnarray}
T^{\mu_1\mu_2\mu_3\mu_4}_{V1}&=&(p_1\cdot p_2 g^{\mu_1\mu_2}-p_2^{\mu_1} p_1^{\mu_2}) (p_1\cdot p_3 g^{\mu_3\mu_4}-p_1^{\mu_3} p_3^{\mu_4}) \;, \\
T^{\mu_1\mu_2\mu_3\mu_4}_{V7}&=&(p_1\cdot p_3 p_2^{\mu_1}-p_1\cdot p_2 p_3^{\mu_1}) (p_2\cdot p_3 g^{\mu_2\mu_3}-p_3^{\mu_2} p_2^{\mu_3}) p_2^{\mu_4} \;,\\
T^{\mu_1\mu_2\mu_3\mu_4}_{V13}&=&(p_1\cdot p_3 g^{\mu_1\mu_2}-p_3^{\mu_1}p_1^{\mu_2})(p_2\cdot p_3g^{\mu_3\mu_4}-p_2^{\mu_3}p_3^{\mu_4})
\nonumber\\
&& +(p_1\cdot p_2 p_3^{\mu_1}-p_1\cdot p_3p_2^{\mu_1})(p_3^{\mu_2}g^{\mu_3\mu_4}-p_3^{\mu_4}g^{\mu_2\mu_3}) \;.
\end{eqnarray}
The corresponding form factors are listed in the Appendix. The remainder gauge structures and form factors can be easily obtained by Bose symmetry, as it is indicated in Table \ref{TABLE8}.

\begin{table}
\caption{\label{TABLE8} Relations dictated by Bose symmetry among the diverse $d_{abc}f^{q}_{V_j} T_{V_j}^{\mu_1\mu_2\mu_3\mu_4}$ terms. }
\begin{ruledtabular}
 \begin{tabular}{|c|c|c|c|}
  $\mathcal{M}_{V\; abc}^{\mu_1\mu_2\mu_3\mu_4}$ & $p_1,\mu_1,a\leftrightarrow p_2,\mu_2,b$ & $p_1,\mu_1,a\leftrightarrow p_3,\mu_3,c$ & $p_2,\mu_2,b\leftrightarrow p_3,\mu_3,c$ \\\hline
   $d_{abc}f^q_{V1}T_{V1}$   &$d_{abc}f^q_{V2}T_{V2}$   &$d_{abc}f^q_{V6}T_{V6}$   &$d_{abc}f^q_{V3}T_{V3}$  \\\hline
   $d_{abc}f^q_{V2}T_{V2}$   &$d_{abc}f^q_{V1}T_{V1}$   &$d_{abc}f^q_{V5}T_{V5}$   &$d_{abc}f^q_{V4}T_{V4}$  \\\hline
   $d_{abc}f^q_{V3}T_{V3}$   &$d_{abc}f^q_{V5}T_{V5}$   &$d_{abc}f^q_{V4}T_{V4}$   &$d_{abc}f^q_{V1}T_{V1}$  \\\hline
   $d_{abc}f^q_{V4}T_{V4}$   &$d_{abc}f^q_{V6}T_{V6}$   &$d_{abc}f^q_{V3}T_{V3}$   &$d_{abc}f^q_{V2}T_{V2}$  \\\hline
   $d_{abc}f^q_{V5}T_{V5}$   &$d_{abc}f^q_{V3}T_{V3}$   &$d_{abc}f^q_{V2}T_{V2}$   &$d_{abc}f^q_{V6}T_{V6}$  \\\hline
   $d_{abc}f^q_{V6}T_{V6}$   &$d_{abc}f^q_{V4}T_{V4}$   &$d_{abc}f^q_{V1}T_{V1}$   &$d_{abc}f^q_{V5}T_{V5}$  \\\hline\hline
   $d_{abc}f^q_{V7}T_{V7}$   &$d_{abc}f^q_{V9}T_{V9}$   &$d_{abc}f^q_{V12}T_{V12}$ &$d_{abc}f^q_{V8}T_{V8}$  \\\hline
   $d_{abc}f^q_{V8}T_{V8}$   &$d_{abc}f^q_{V10}T_{V10}$ &$d_{abc}f^q_{V11}T_{V11}$ &$d_{abc}f^q_{V7}T_{V7}$  \\\hline
   $d_{abc}f^q_{V9}T_{V9}$   &$d_{abc}f^q_{V7}T_{V7}$   &$d_{abc}f^q_{V10}T_{V10}$ &$d_{abc}f^q_{V11}T_{V11}$  \\\hline
   $d_{abc}f^q_{V10}T_{V10}$ &$d_{abc}f^q_{V8}T_{V8}$   &$d_{abc}f^q_{V9}T_{V9}$   &$d_{abc}f^q_{V12}T_{V12}$  \\\hline
   $d_{abc}f^q_{V11}T_{V11}$ &$d_{abc}f^q_{V12}T_{V12}$ &$d_{abc}f^q_{V8}T_{V8}$   &$d_{abc}f^q_{V9}T_{V9}$    \\\hline
   $d_{abc}f^q_{V12}T_{V12}$ &$d_{abc}f^q_{V11}T_{V11}$ &$d_{abc}f^q_{V7}T_{V7}$   &$d_{abc}f^q_{V10}T_{V10}$  \\\hline\hline
   $d_{abc}f^q_{V13}T_{V13}$ &$d_{abc}f^q_{V14}T_{V14}$ &$d_{abc}f^q_{V17}T_{V17}$ &$d_{abc}f^q_{V16}T_{V16}$  \\\hline
   $d_{abc}f^q_{V14}T_{V14}$ &$d_{abc}f^q_{V13}T_{V13}$ &$d_{abc}f^q_{V18}T_{V18}$ &$d_{abc}f^q_{V15}T_{V15}$  \\\hline
   $d_{abc}f^q_{V15}T_{V15}$ &$d_{abc}f^q_{V17}T_{V17}$ &$d_{abc}f^q_{V16}T_{V16}$ &$d_{abc}f^q_{V14}T_{V14}$  \\\hline
   $d_{abc}f^q_{V16}T_{V16}$ &$d_{abc}f^q_{V18}T_{V18}$ &$d_{abc}f^q_{V15}T_{V15}$ &$d_{abc}f^q_{V13}T_{V13}$  \\\hline
   $d_{abc}f^q_{V17}T_{V17}$ &$d_{abc}f^q_{V15}T_{V15}$ &$d_{abc}f^q_{V13}T_{V13}$ &$d_{abc}f^q_{V18}T_{V18}$  \\\hline
   $d_{abc}f^q_{V18}T_{V18}$ &$d_{abc}f^q_{V16}T_{V16}$ &$d_{abc}f^q_{V14}T_{V14}$ &$d_{abc}f^q_{V17}T_{V17}$
\end{tabular}
\end{ruledtabular}
\end{table}

We now turn to discuss the mathematical structure of the axial vector amplitude. As already mentioned, this amplitude receives contributions from both box and triangle diagrams, in contrast with the vector amplitude to which contribute only the box diagrams. While the contributions of both box and triangle graphs satisfy separately the Bose symmetry, one needs to sum over both type of contributions in order to obtain invariance under the color group. Due to this, it is more difficult to conciliate both class of symmetries in order to write compact expression, as in the vector case. So, while a judicious use of the Schouthen's identity \cite{SI} allows us to write the amplitude in terms of 21 Lorentz tensor gauge structures, explicit Bose symmetry is sacrificed. However, we have find that if the number of gauge structures is enhanced to 24, both gauge and Bose symmetries can be maintained in a manifest way. In this basis, the axial vector amplitude can be written as:
\begin{equation}
\mathcal{M}_{A\;abc}^{\mu_1\mu_2\mu_3\mu_4}=
g_{AV}^{q}f_{abc}\left(-\frac{ig_s^3g_VN_C}{4\pi^2}\right)
\sum_{j=1}^{24} f^{q}_{A_j} T_{A_j}^{\mu_1\mu_2\mu_3\mu_4} \;,
\end{equation}
where the $f^{q}_{A_j}$ coefficients are Lorentz scalars form factors, whereas and the $T_{A_j}^{\mu_1\mu_2\mu_3\mu_4}$ tensors are gauge structures satisfying the transversality conditions:
\begin{equation}
p_{j\mu_j}T_{A_j}^{\mu_1\mu_2\mu_3\mu_4}=0 \; , \; j=1,2,3\; .
\end{equation}
In this extended basis, the axial vector amplitude can be written in terms of compact expressions. As it occurs for the vector amplitude, in this case the set 24 gauge structures, together with their 24 associated form factors, can be classified into 4 subsets, each composed of 6 elements, all them related through Bose symmetry. In this way, it is only necessary to write one representative element of each subset. Accordingly, we have chosen the following representative gauge structures:
\begin{eqnarray}
T_{A 1}^{\mu_1 \mu_2 \mu_3 \mu_4} & = & \epsilon^{\mu_3 \mu_4 p_1 p_3} (p_2^{\mu_1} p_1^{\mu_2} - p_1\cdot p_2 g^{\mu_1 \mu_2}) \ , \\
T_{A 7}^{\mu_1 \mu_2 \mu_3 \mu_4} & = & (p_3^{\mu_1} \epsilon^{\mu_3 \mu_4 p_1 p_3} - p_1\cdot p_3 \epsilon^{\mu_1 \mu_3 \mu_4 p_3})(p_1\cdot p_2 p_3^{\mu_2} - p_2\cdot p_3 p_1^{\mu_2}) \ , \\
T_{A 13}^{\mu_1 \mu_2 \mu_3 \mu_4} & = & \epsilon^{\mu_1 \mu_3 \mu_4 p_3} (p_2\cdot p_3 p_1^{\mu_2} - p_1\cdot p_2 p_3^{\mu_2}) + \epsilon^{\mu_3 \mu_4 p_1 p_3} (p_2^{\mu_1} p_3^{\mu_2} - p_2\cdot p_3 g^{\mu_1 \mu_2}) \ , \\
T_{A 19}^{\mu_1 \mu_2 \mu_3 \mu_4} & = & p_1\cdot p_2 (p_3^{\mu_2} \epsilon^{\mu_1 \mu_3 \mu_4 p_2} - p_2^{\mu_3} \epsilon^{\mu_1 \mu_2 \mu_4 p_3} - g^{\mu_2 \mu_3} \epsilon^{\mu_1 \mu_4 p_2 p_3} - p_2\cdot p_3 \epsilon^{\mu_1 \mu_2 \mu_3 \mu_4}) \\ & & + p_2^{\mu_1} (p_2^{\mu_3} \epsilon^{\mu_2 \mu_4 p_1 p_3} - p_3^{\mu_2} \epsilon^{\mu_3 \mu_4 p_1 p_2} - g^{\mu_2 \mu_3} \epsilon^{\mu_4 p_1 p_2 p_3} - p_2\cdot p_3 \epsilon^{\mu_2 \mu_3 \mu_4 p_1}) \ .
\end{eqnarray}
The corresponding form factors are listed in the appendix. Starting from these representative form factors and gauge structures, it is easy to construct explicitly the remainder ones, as it is illustrated in Table \ref{TABLE9}.

\begin{table}
\caption{\label{TABLE9} Relations dictated by Bose symmetry among the diverse $f_{abc}f^{q}_{A_j} T_{A_j}^{\mu_1\mu_2\mu_3\mu_4}$ terms. }
\begin{ruledtabular}
 \begin{tabular}{|c|c|c|c|}
$\mathcal{M}_{A\ abc}^{\mu_1 \mu_2 \mu_3 \mu_4}$ & $p_1, \mu_1, a \leftrightarrow p_2, \mu_2, b$ & $p_1, \mu_1, a \leftrightarrow p_3, \mu_3, c$ & $p_2, \mu_2, b \leftrightarrow p_3, \mu_3, c$ \\
\hline
$f_{abc} f^q_{A1} T_{A1}$ & $f_{abc} f^q_{A2} T_{A2}$ & $f_{abc} f^q_{A6} T_{A6}$ & $f_{abc} f^q_{A3} T_{A3}$ \\
\hline
$f_{abc} f^q_{A2} T_{A2}$ & $f_{abc} f^q_{A1} T_{A1}$ & $f_{abc} f^q_{A5} T_{A5}$ & $f_{abc} f^q_{A4} T_{A4}$ \\
\hline
$f_{abc} f^q_{A3} T_{A3}$ & $f_{abc} f^q_{A5} T_{A5}$ & $f_{abc} f^q_{A4} T_{A4}$ & $f_{abc} f^q_{A1} T_{A1}$ \\
\hline
$f_{abc} f^q_{A4} T_{A4}$ & $f_{abc} f^q_{A6} T_{A6}$ & $f_{abc} f^q_{A3} T_{A3}$ & $f_{abc} f^q_{A2} T_{A2}$ \\
\hline
$f_{abc} f^q_{A5} T_{A5}$ & $f_{abc} f^q_{A3} T_{A3}$ & $f_{abc} f^q_{A2} T_{A2}$ & $f_{abc} f^q_{A6} T_{A6}$ \\
\hline
$f_{abc} f^q_{A6} T_{A6}$ & $f_{abc} f^q_{A4} T_{A4}$ & $f_{abc} f^q_{A1} T_{A1}$ & $f_{abc} f^q_{A5} T_{A5}$ \\
\hline\hline
$f_{abc} f^q_{A7} T_{A7}$ & $f_{abc} f^q_{A8} T_{A8}$ & $f_{abc} f^q_{A12} T_{A12}$ & $f_{abc} f^q_{A9} T_{A9}$ \\
\hline
$f_{abc} f^q_{A8} T_{A8}$ & $f_{abc} f^q_{A7} T_{A7}$ & $f_{abc} f^q_{A11} T_{A11}$ & $f_{abc} f^q_{A10} T_{A10}$  \\
\hline
$f_{abc} f^q_{A9} T_{A9}$ & $f_{abc} f^q_{A11} T_{A11}$ & $f_{abc} f^q_{A10} T_{A10}$ & $f_{abc} f^q_{A7} T_{A7}$ \\
\hline
$f_{abc} f^q_{A10} T_{A10}$ & $f_{abc} f^q_{A12} T_{A12}$ & $f_{abc} f^q_{A9} T_{A9}$ & $f_{abc} f^q_{A8} T_{A8}$ \\
\hline
$f_{abc} f^q_{A11} T_{A11}$ & $f_{abc} f^q_{A9} T_{A9}$ & $f_{abc} f^q_{A8} T_{A8}$ & $f_{abc} f^q_{A12} T_{A12}$ \\
\hline
$f_{abc} f^q_{A12} T_{A12}$ & $f_{abc} f^q_{A10} T_{A10}$ & $f_{abc} f^q_{A7} T_{A7}$ & $f_{abc} f^q_{A11} T_{A11}$ \\
\hline\hline
$f_{abc} f^q_{A13} T_{A13}$ & $f_{abc} f^q_{A14} T_{A14}$ & $f_{abc} f^q_{A17} T_{A17}$ & $f_{abc} f^q_{A16} T_{A16}$ \\
\hline
$f_{abc} f^q_{A14} T_{A14}$ & $f_{abc} f^q_{A13} T_{A13}$ & $f_{abc} f^q_{A18} T_{A18}$ & $f_{abc} f^q_{A15} T_{A15}$ \\
\hline
$f_{abc} f^q_{A15} T_{A15}$ & $f_{abc} f^q_{A17} T_{A17}$ & $f_{abc} f^q_{A16} T_{A16}$ & $f_{abc} f^q_{A14} T_{A14}$ \\
\hline
$f_{abc} f^q_{A16} T_{A16}$ & $f_{abc} f^q_{A18} T_{A18}$ & $f_{abc} f^q_{A15} T_{A15}$ & $f_{abc} f^q_{A13} T_{A13}$ \\
\hline
$f_{abc} f^q_{A17} T_{A17}$ & $f_{abc} f^q_{A15} T_{A15}$ & $f_{abc} f^q_{A13} T_{A13}$ & $f_{abc} f^q_{A18} T_{A18}$ \\
\hline
$f_{abc} f^q_{A18} T_{A18}$ & $f_{abc} f^q_{A16} T_{A16}$ & $f_{abc} f^q_{A14} T_{A14}$ & $f_{abc} f^q_{A17} T_{A17}$ \\
\hline\hline
$f_{abc} f^q_{A19} T_{A19}$ & $f_{abc} f^q_{A21} T_{A21}$ & $f_{abc} f^q_{A24} T_{A24}$ & $f_{abc} f^q_{A20} T_{A20}$ \\
\hline
$f_{abc} f^q_{A20} T_{A20}$ & $f_{abc} f^q_{A22} T_{A22}$ & $f_{abc} f^q_{A23} T_{A23}$ & $f_{abc} f^q_{A19} T_{A19}$ \\
\hline
$f_{abc} f^q_{A21} T_{A21}$ & $f_{abc} f^q_{A19} T_{A19}$ & $f_{abc} f^q_{A22} T_{A22}$ & $f_{abc} f^q_{A23} T_{A23}$ \\
\hline
$f_{abc} f^q_{A22} T_{A22}$ & $f_{abc} f^q_{A20} T_{A20}$ & $f_{abc} f^q_{A21} T_{A21}$ & $f_{abc} f^q_{A24} T_{A24}$ \\
\hline
$f_{abc} f^q_{A23} T_{A23}$ & $f_{abc} f^q_{A24} T_{A24}$ & $f_{abc} f^q_{A20} T_{A20}$ & $f_{abc} f^q_{A21} T_{A21}$ \\
\hline
$f_{abc} f^q_{A24} T_{A24}$ & $f_{abc} f^q_{A23} T_{A23}$ & $f_{abc} f^q_{A19} T_{A19}$ & $f_{abc} f^q_{A22} T_{A22}$ \\
\end{tabular}
\end{ruledtabular}
\end{table}

%%%%%%%%%%%%%%%%%%%%%%%%%%%%%%%%%%%%%%%%%%%%%%%%%%%%%%FIGURE 1
%%%%%%%%%%%%%%%%%%%%%%%%%%%%%%%%%%%%%%%%%%%%%%%%%%%%%%%%%%%%

%%%%%%%%%%%%%%%%%%%%%%%%%%%%%%%%%%%%%%%%%%%%%%%%%%%%%%FIGURE 2
%%%%%%%%%%%%%%%%%%%%%%%%%%%%%%%%%%%%%%%%%%%%%%%%%%%%%%%%%%%%

\section{Results and discussion}
\label{rd}In this section, we discuss our results for the branching ratios of the $Z\to ggg$~\cite{PTT} and $Z'\to ggg$ decays. The expression for the decay width of the $V\to ggg$ transition can be write in a generic way as follows:
\begin{eqnarray}
  \Gamma(V\rightarrow ggg) &=& \frac{m_V}{3!\;256\; \pi^3}
  \int_0^1\int_{1-x}^1 |\mathcal{M}|^2 dy dx \nonumber\\
  &=& \frac{\alpha_s^3(m_V)\alpha N_C^2 m_V }{384\;\pi^3c_W^2s_W^2}
  \int_0^1\int_{1-x}^1
  \sum_{q,q'} \Bigg[\frac{40}{3}\;g_{VV}^{q}g_{VV}^{q'}\Bigg(\frac{1}{3}\sum_{\lambda_1,\lambda_2, \lambda_3, \lambda_4}\mathcal{V}_{q}\mathcal{V}_{q'}^*\Bigg)\nonumber \\
  &&+24\;g_{AV}^{q}g_{AV}^{q'}\Bigg(\frac{1}{3}\sum_{\lambda_1,\lambda_2, \lambda_3, \lambda_4}\mathcal{A}_{q}\mathcal{A}_{q'}^* \Bigg)\Bigg] \;dy dx
\end{eqnarray}
where the sums in $\lambda_i$ represent the bosons polarization sums. The last expression was obtained after using the following definition
\begin{equation}
\mathcal{M}_{V\to ggg}=g^{q}_{VV}d_{abc}\Big(-\frac{ig^3_sg_VN_C}{4\pi^2}\Big)\mathcal{V}_{q}+
g^{q}_{AV}f_{abc}\Big(-\frac{ig^3_sg_VN_C}{4\pi^2}\Big)\mathcal{A}_{q}\;,
\end{equation}
with
\begin{eqnarray}
\mathcal{V}_{q}&=&\sum_{j=1}^{18} f_{V_j}^{q} T_{V_j}^{\mu_1\mu_2\mu_3\mu_4}
\epsilon_{\mu_1}^{*\,a}(p_1,\lambda_1)\epsilon_{\mu_2}^{*\,b}(p_2,\lambda_2)
\epsilon_{\mu_3}^{*\,c}(p_3,\lambda_3)\epsilon_{\mu_4}(p_4,\lambda_4),\\
\mathcal{A}_{q}&=&\sum_{j=1}^{24} f_{A_j}^{q} T_{A_j}^{\mu_1\mu_2\mu_3\mu_4}
\epsilon_{\mu_1}^{*\,a}(p_1,\lambda_1)\epsilon_{\mu_2}^{*\,b}(p_2,\lambda_2)
\epsilon_{\mu_3}^{*\,c}(p_3,\lambda_3)\epsilon_{\mu_4}(p_4,\lambda_4) \;.
\end{eqnarray}
The phase space dimensionless variables $x$ and $y$ are defined by
\begin{equation}
x=\frac{2p^0_1}{m_V}\; , \; y=\frac{2p^0_2}{m_V}\; , \;  z=\frac{2p^0_3}{m_V}\; ,\\
\end{equation}
which satisfy the relation $x+y+z=2$. In terms of these variables, the scalar products $p_i\cdot p_j$ are given by
\begin{eqnarray}
&&p_1\cdot p_2=\frac{m_V^2}{2}(x+y-1)\;,\\
&&p_1\cdot p_3=\frac{m_V^2}{2}(1-y)\;,\\
&&p_2\cdot p_3=\frac{m_V^2}{2}(1-x)\;.
\end{eqnarray}
The definition domain of these variables is: $0\leq x \leq 1 $ and $ 1-x \leq y \leq 1 $. We now are ready to present numerical results. In obtaining these numerical results, the Passarino-Veltman scalar functions were evaluated numerically using FF routines \cite{FF}.

\subsection{Decay $Z\to ggg$}
\begin{figure}
\centering
\includegraphics[width=3.5in]{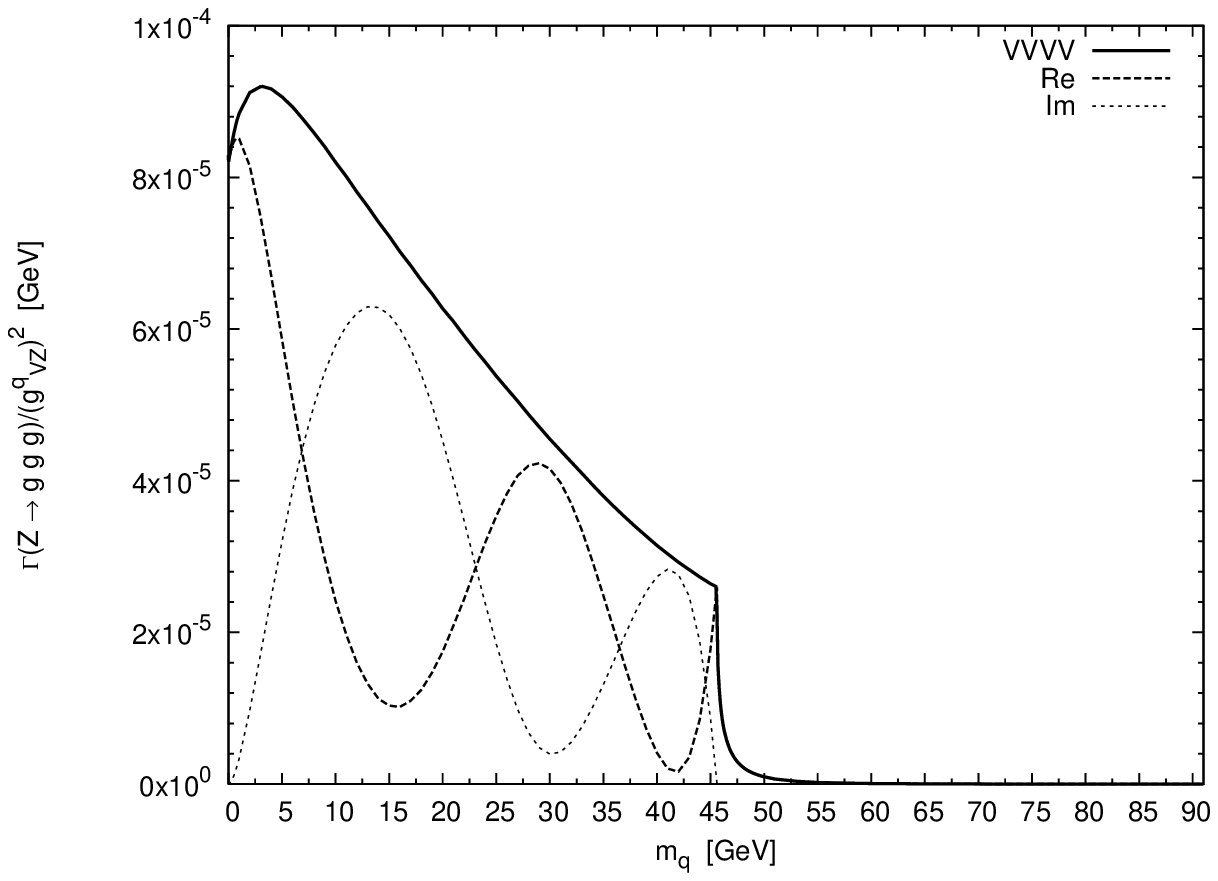}
\includegraphics[width=3.5in]{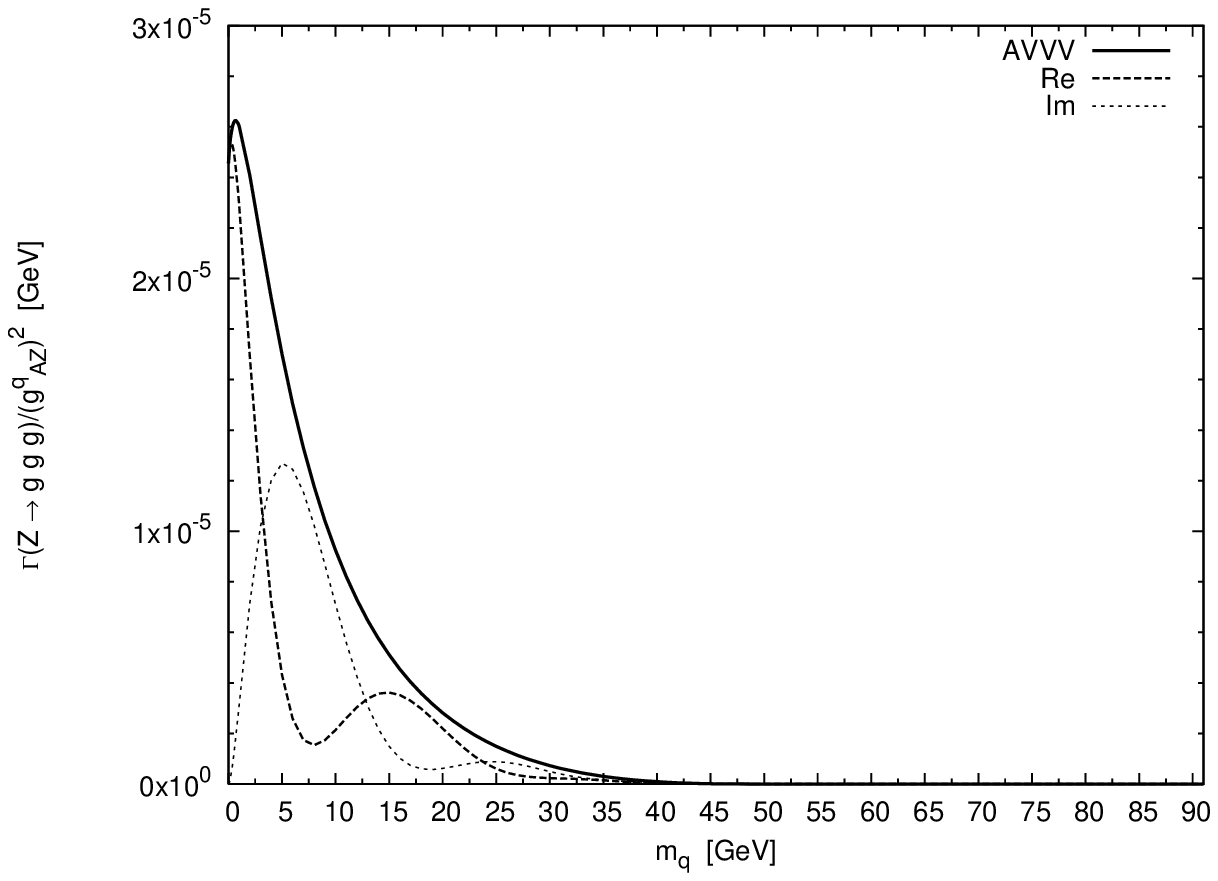}
\caption{\label{D}Decoupling of the vector and the axial vector amplitudes of the $Z\to ggg$ decay when considered as a function of the quark mass. The behavior of both the real and imaginary parts of the amplitudes are shown. }
\end{figure}

\begin{figure}
\centering
\includegraphics[width=4.0in]{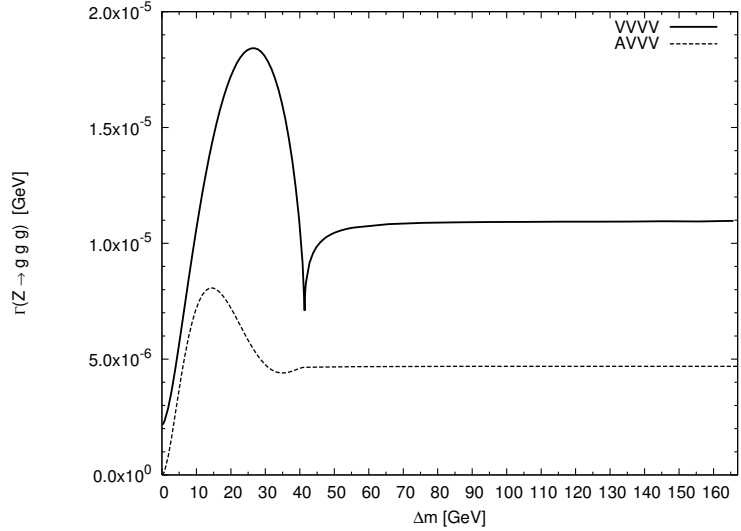}
\caption{\label{NDF}Nondecoupling of vector and axial vector amplitudes of the $Z\to ggg$ decay as a function of the mass difference of the members of the doublet: $\Delta m=m_u-m_d$. The graphic shown corresponds to the case $m_d=0$.}
\end{figure}

In the minimal $331$ model, the contribution to the decay width of the $Z\to ggg$ transition can be written as the sum of three partial widths:
\begin{equation}
\Gamma (Z\to ggg)=\Gamma_{q_i}+\Gamma_{Q_i}+\Gamma_{q_i-Q_i}\; ,
\end{equation}
were $\Gamma_{q_i}$, $\Gamma_{Q_i}$, and $\Gamma_{q_i-Q_i}$ are the contributions of the SM quarks, the exotic quarks, and the interference between these contributions, respectively. Before presenting the numerical values for these quantities, let us to present a brief discussion about the decoupling nature of the vector and the axial vector amplitudes when considered as a function of the quark mass. In Fig.~\ref{D} the behavior of the vector amplitude (VVVV) (left) and the axial vector amplitude (AVVV) (right) are shown  as a function of the quark mass. The behavior is shown for the bare amplitudes $\Gamma(Z\to ggg)/(g^{q}_{VZ})^2$ and $\Gamma(Z\to ggg)/(g^{q}_{AZ})^2$. It can be seen from this figure that these amplitudes vanish in the heavy mass limit, which shows their decoupling nature. The behavior of the real and imaginary parts of the amplitudes are shown too. From this figure, it can be appreciated that the width decay reach its maximum value for a quark mass of about $m_q=3.2$ GeV and immediately drop to a negligible value. As we will se below, the vector amplitude is dominated by the bottom quark. As it can be appreciated from Fig.~\ref{D}, the bare axial vector amplitude reach its maximum value for $m_q=0.67$ GeV. Since the axial vector couplings of $Z$ to up and down quarks are equal in magnitude but have opposite signs, there is no contribution in the degenerate case, but a maximum contribution is found for the highest mass difference of the members of a family. Consequently, the dominant contribution to this amplitude arise from the third family. Indeed, both the vector and axial vector amplitudes present a nondecoupling behavior when considered as a function of the mass difference between the members of a family, as they tend to a finite nonzero value for a large mass difference. This behavior, which nicely reproduces the results given in ref.~\cite{TopE}, is shown in Fig.~\ref{NDF}.

We now proceed to present numerical results. We will use the following values for the various parameters appearing in the amplitudes~\cite{PDG}:
$m_Z=91_\cdot1876\;\mathrm{GeV}$, $ m_u=0_\cdot00255\;\mathrm{GeV}$, $ m_d=0_\cdot00504\;\mathrm{GeV}$, $ m_s=0_\cdot104\;\mathrm{GeV}$, $ m_c=1_\cdot27\;\mathrm{GeV}$, $ m_b=4_\cdot2\;\mathrm{GeV}$, $m_t=171_\cdot2\;\mathrm{GeV}$, $s_W^2=0_\cdot23119$, $\alpha_s(m_Z)=0_\cdot1176$, and $\alpha(m_Z)=1/128$. Regarding to the masses of the exotic quarks, the lower bound $m_Q>240$ GeV was derived from the search for supersymmetry at the Tevatron and would reach the level of 320 at run 2~\cite{EQB}. In Ref.~\cite{EQB1} the production of exotic quarks at THERA and LHC via $E_6$ theories has been studied, they have found that exotic quarks mass can be high as 450 GeV and 1.2 TeV. It is then reasonable to consider the range 500 GeV$\,\leq m_Q \leq\,$700 GeV for our numerical analysis. In this scenario we will consider that $m_{D,S,T}\,=\,500$ GeV. With these values, one obtains
\begin{eqnarray}
\Gamma_{q_i}&=&3.49\times 10^{-5}\; \mathrm{GeV}\; ,\\
\Gamma_{Q_i}&\sim &10^{-12} \; \mathrm{GeV}\; ,\\
\Gamma_{q_i-Q_i}&\sim & 10^{-10}\; \mathrm{GeV}\; .
\end{eqnarray}
From these results, it is clear that the exotic quark contribution is absolutely marginal. As far as the contribution of the known quark is concerned, in Table.~\ref{TABLE2} a more detailed information is presented. From this table, it can be appreciated that both the vector amplitude and the axial vector amplitude are essentially determined by the third family and that the latter is almost one order of magnitude lower than the former. All our result are in perfect agreement with those given in the literature, especially with those presented in ref.~\cite{TopE}.

\begin{table}
\caption{\label{TABLE2} Family contribution to the $\Gamma(Z\to ggg)$ decay in the standard model. Here, $\Gamma^{VI}$ and $\Gamma^{AI}$ represent the interference effect induced by the three families into the vector and axial vector width decays, respectively.}
\begin{ruledtabular}
\begin{center}
\begin{tabular}{|c|r|c|c|c|c|c|c|}
  Family & $\Gamma^V$ [GeV] & $\Gamma^A$ [GeV] & $\Gamma^{VI}$ [GeV] & $\Gamma^{AI}$ [GeV] & $\Gamma_{q_i}$ [GeV] & $\Gamma_{Q_i}$ [GeV] & $\Gamma_{q_i-Q_i}$ [GeV]\\
  \hline
  $u,d$ & $1.95\times 10^{-6}$  & $\sim 10^{-11}$       & - & - & - & - & - \\
 \hline
  $c,s$ & $2.21\times 10^{-6}$  & $1.5\times 10^{-6}$  & - & - & - & - & - \\
  \hline
  $t,b$ & $1.09\times 10^{-5}$  & $4.69\times 10^{-6}$ & - & - & - & - & - \\
  \hline
  Total & $1.51\times 10^{-5}$  & $6.19\times 10^{-6}$ & $1.66\times 10^{-5}$ & $-3.03\times 10^{-6}$ & $3.49\times 10^{-5}$ & $\sim 10^{-12}$ & $\sim 10^{-10}$
\end{tabular}
\end{center}
\end{ruledtabular}
\end{table}

Finally, the branching ratio for the $Z\to ggg$ decay in the minimal $331$ model is given by
\begin{equation}
Br(Z\to ggg)=1.4 \times 10^{-5}\; ,
\end{equation}
which is determined essentially by the third family of quarks, as the contribution of the exotic quark is negligible.

%%%%%%%%%%%%%%%%%%%%%%%%%%%%%%%%%%%%%%%%%%%%%%%%%%%%%%FIGURE 3
%%%%%%%%%%%%%%%%%%%%%%%%%%%%%%%%%%%%%%%%%%%%%%%%%%%%%%%%%%%%

%%%%%%%%%%%%%%%%%%%%%%%%%%%%%%%%%%%%%%%%%%%%%%%%%%%%%%FIGURE 4
%%%%%%%%%%%%%%%%%%%%%%%%%%%%%%%%%%%%%%%%%%%%%%%%%%%%%%%%%%%%

\subsection{Decay $Z'\to ggg$}
\begin{figure}
\centering
\includegraphics[width=3.5in]{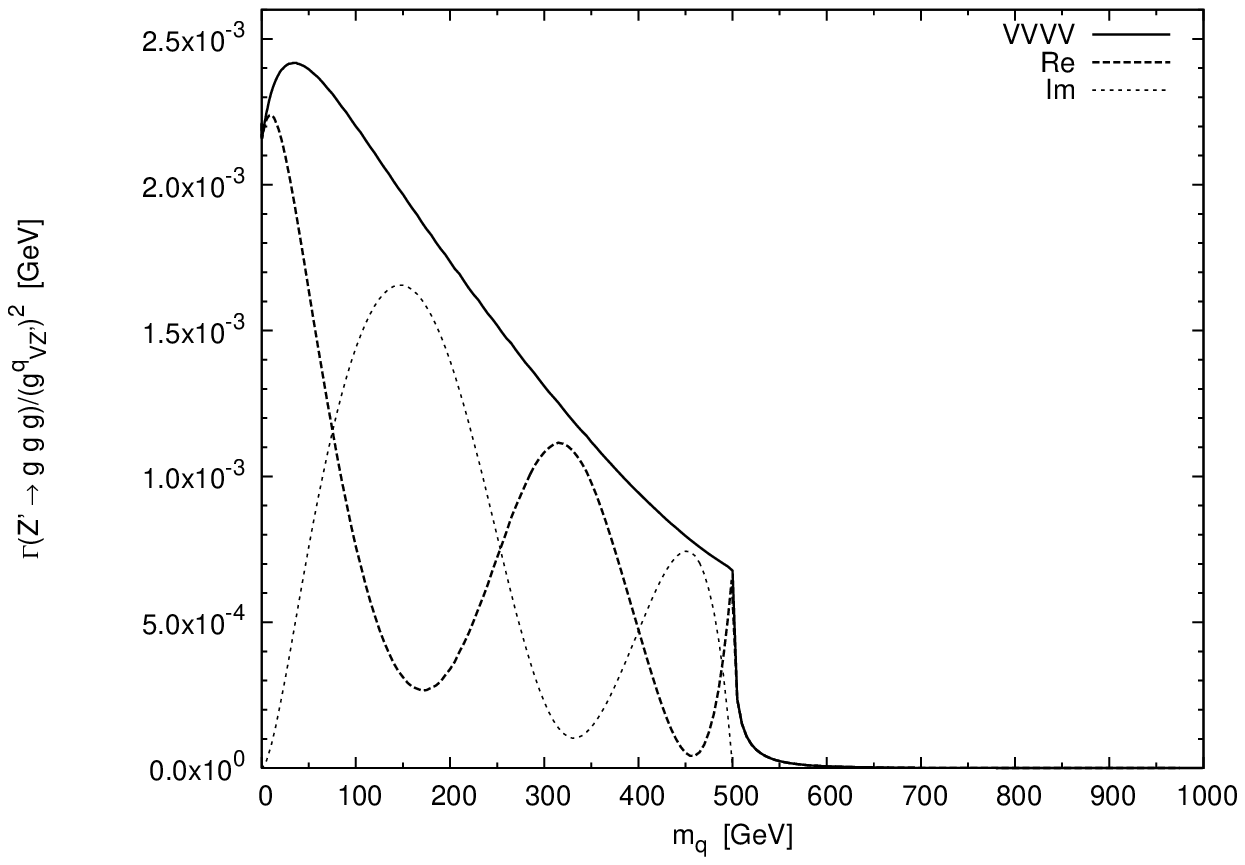}
\includegraphics[width=3.5in]{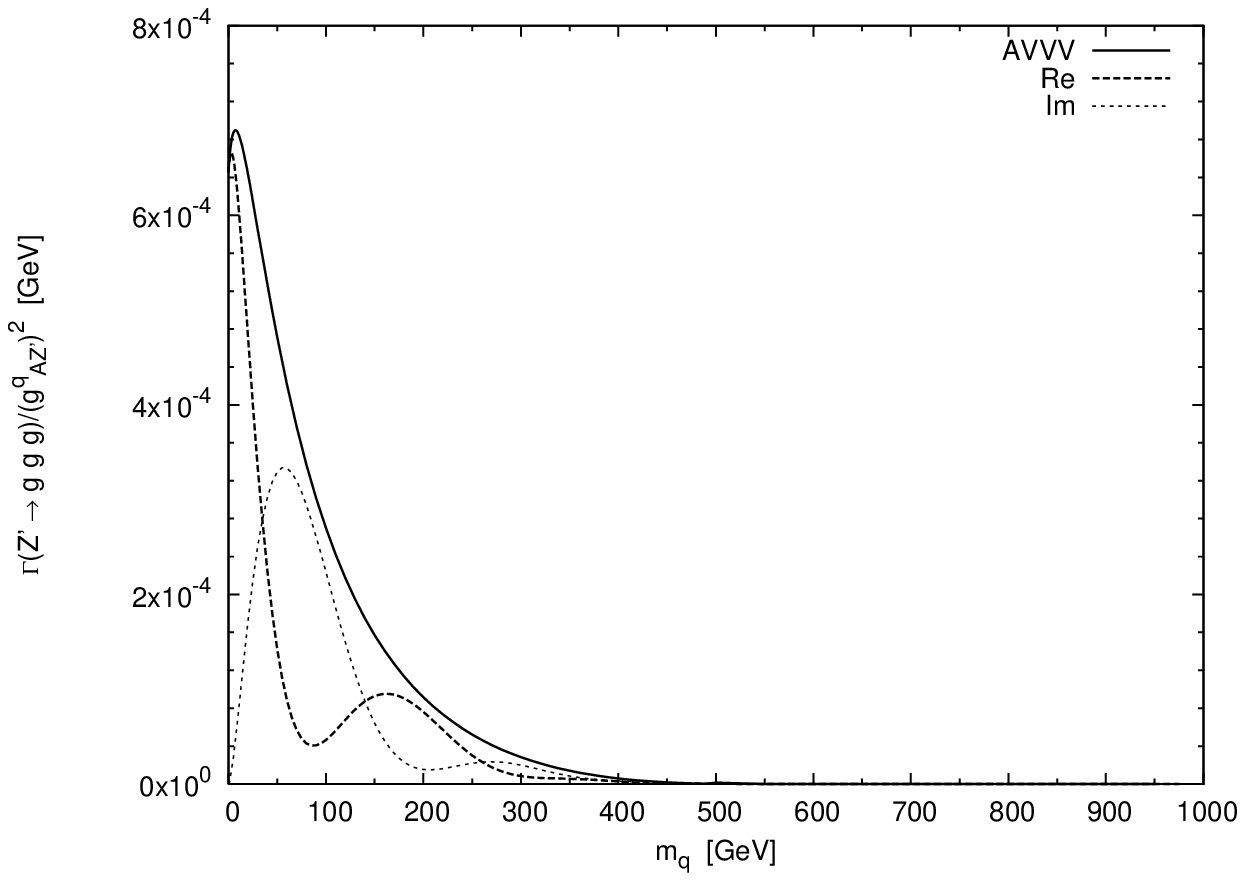}
\caption{\label{DZP}Decoupling of the vector and the axial vector amplitudes of the $Z'\to ggg$ decay when considered as a function of the quark mass. The behavior of both the real and imaginary parts of the amplitudes are shown. }
\end{figure}

\begin{figure}
\centering
\includegraphics[width=3.5in]{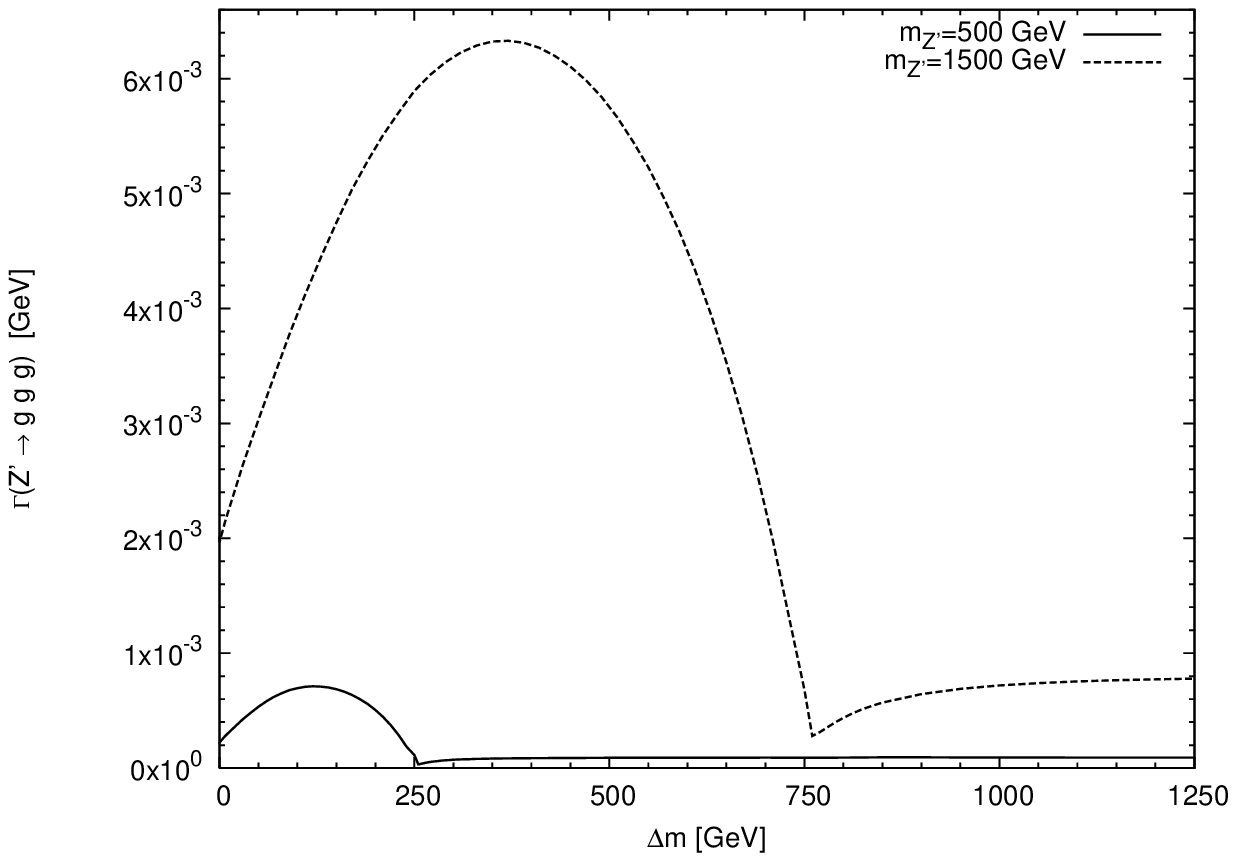}
\includegraphics[width=3.5in]{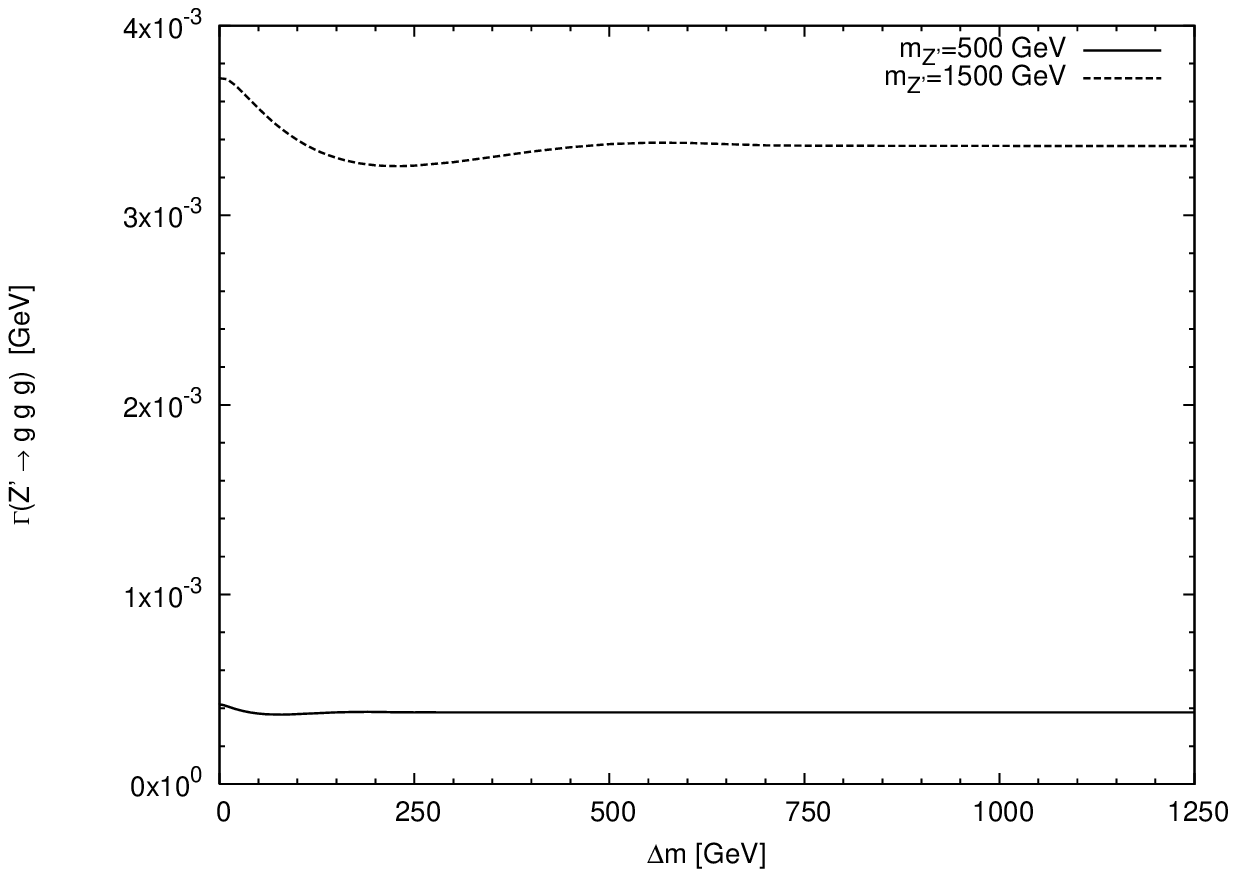}
\caption{\label{NDZPF1}Nondecoupling behavior of the vector (left) and the axial vector (right) amplitudes of the $Z'\to ggg$ decay when considered as a function of the mass difference of the members of the doublet of the first family. The behavior for the second family is identic. }
\end{figure}

\begin{figure}
\centering
\includegraphics[width=3.5in]{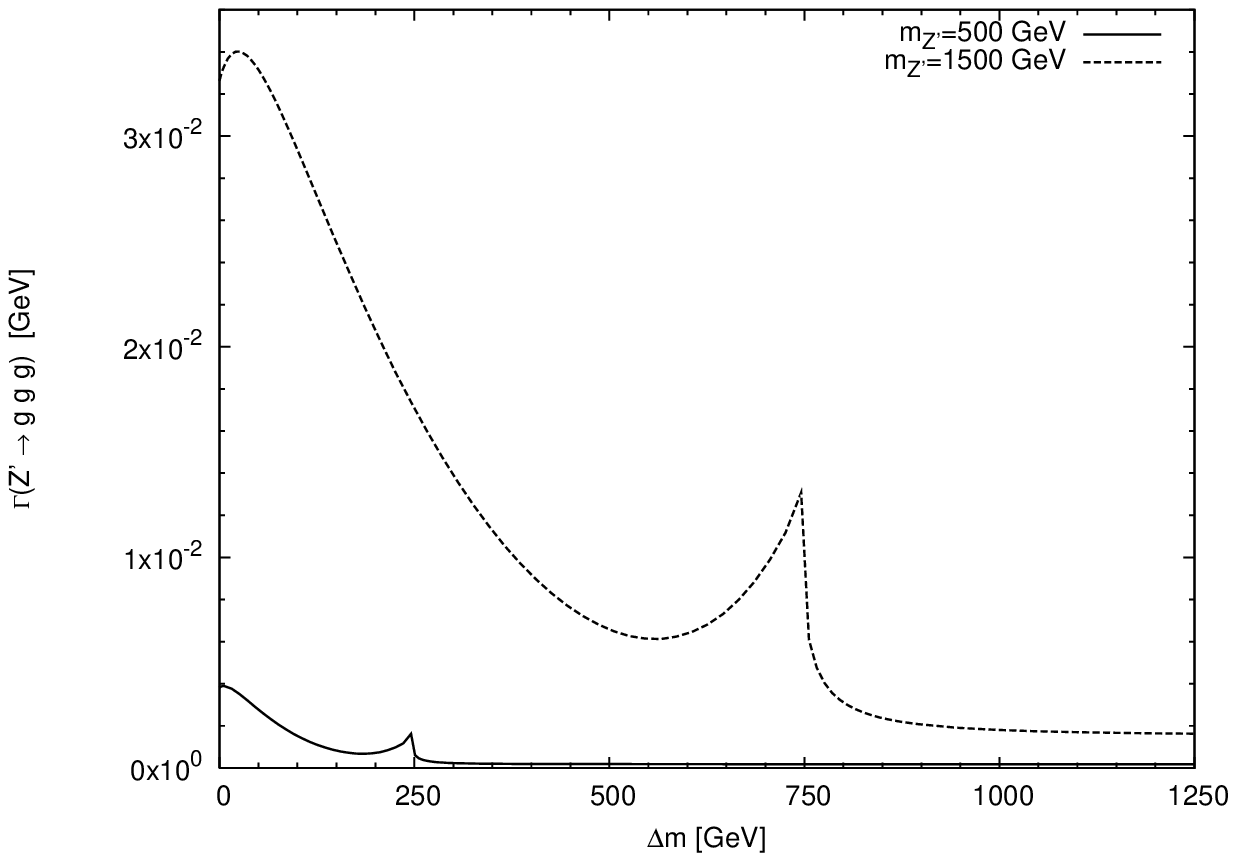}
\includegraphics[width=3.5in]{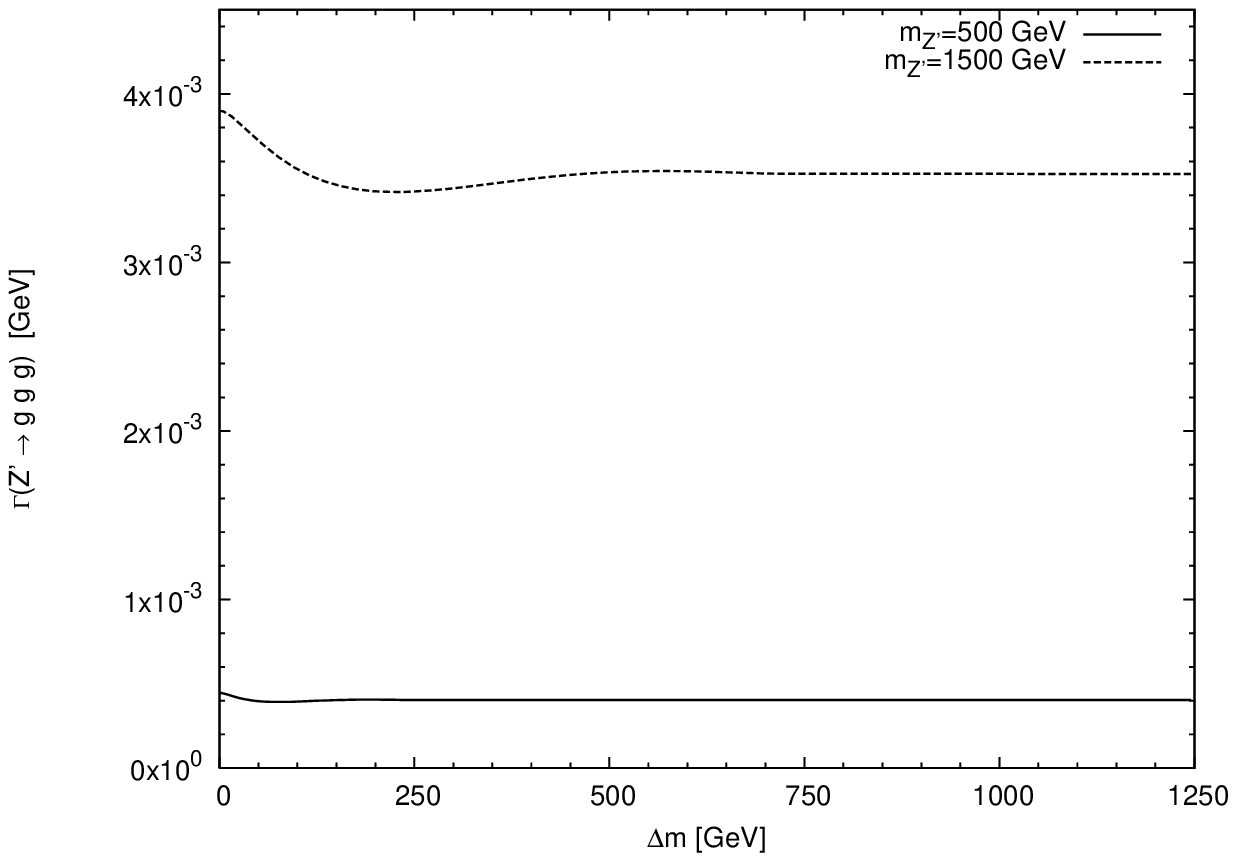}
\caption{\label{NDZPF3}Nondecoupling behavior of the vector (left) and the axial vector (right) amplitudes of the $Z'\to ggg$ decay when considered as a function of the mass difference of the members of the doublet of the third family.}
\end{figure}

We now turn to present numerical results for the $Z'\to ggg$ decay. Although the mathematical structure of the decay width is identical to the one associated with the $Z\to ggg$ decay, its numerical behavior present some differences due to the fact that the $331$ model treats the third family differently to the other two. As it can be appreciated from Table~\ref{TABLE}, the main differences between the $Z'\bar{q}q$  and $Z\bar{q}q$ couplings are the following: 1) the vector ($g^{q}_{VZ'}$) and axial vector ($g^{q}_{AZ'}$) couplings, which are about one order of magnitude larger than the respective couplings of the $Z$ boson. As we will se below, these facts lead to partial decay widths larger than those associated with the $Z$ boson. 2) the axial vector couplings of $Z'$ to the members of a doublet are not the negative one of the other, as it occurs for the case of the standard $Z$ boson. 3) the $Z'$ coupling to the third family differs from its couplings to the first two, which are a replica one of the other. As in the case of the $Z\to ggg$ decay, we express the decay width into three contributions:
\begin{equation}
\Gamma (Z'\to ggg)=\Gamma_{q_i}+\Gamma_{Q_i}+\Gamma_{q_i-Q_i}\; ,
\end{equation}
were $\Gamma_{q_i}$, $\Gamma_{Q_i}$, and $\Gamma_{q_i-Q_i}$ are the contributions of the SM quarks, the exotic quarks, and the interference between these contributions, respectively. As far as the $Z'$ boson mass is concerned, although it is not possible to obtain model-independent bounds, current limits from precision experiments implies that $m_{Z'}\gtrsim$ 500 GeV~\cite{T1}. Similar bounds was obtained in Ref.~\cite{Zpmass1} from both 331 minimal model and 331 model with right-handed neutrinos. In Ref.~\cite{Zpmass2} a bound for $Z'$ mass about of the order of 300 GeV has been obtained from 331 models at electroweak scale. Studies in the context of 331 models predicts lower bounds greater than 1.5 TeV~\cite{Zpmass3}. In addition, model-dependent upper bounds of $Z'$ mass are imposed too by means of the Landau pole in the context of a perturbative treatment of the model \cite{LP}, where such bounds are usually estimated around of 3 TeV. Therefore, we have considered four scenarios corresponding to $m_{Z'}=500$, $1000$, $2000$, and $3000$ GeV for decoupling analysis, to which it is found a maximum value for the vector amplitude in values of quark masses of $m_q=18$, $35$, $71$, and $107$ GeV, respectively. A similar behavior is observed for the axial vector contribution when considered as a function of the  quark mass. In Fig.~\ref{DZP} the decoupling nature of the partial vector and axial vector decay widths are shown as a function of the quark mass for the case $m_{Z'}=1000 $ GeV. The nondecoupling nature of both the vector and axial vector contributions when considered as a function of the mass difference between the members of a doublet is shown in Figs. \ref{NDZPF1} and \ref{NDZPF3}. It is interesting to compare these figures with figure \ref{NDF}, from which a very different behavior on the nondecoupling nature of the amplitudes can be appreciated.

%%%%%%%%%%%%%%%%%%%%%%%%%%%%%%%%%%%%%%%%%%%%%%%%%%%%%%FIGURE 5
%%%%%%%%%%%%%%%%%%%%%%%%%%%%%%%%%%%%%%%%%%%%%%%%%%%%%%%%%%%%

%%%%%%%%%%%%%%%%%%%%%%%%%%%%%%%%%%%%%%%%%%%%%%%%%%%%%%FIGURE 6
%%%%%%%%%%%%%%%%%%%%%%%%%%%%%%%%%%%%%%%%%%%%%%%%%%%%%%%%%%%%

%%%%%%%%%%%%%%%%%%%%%%%%%%%%%%%%%%%%%%%%%%%%%%%%%%%%%%FIGURE 7
%%%%%%%%%%%%%%%%%%%%%%%%%%%%%%%%%%%%%%%%%%%%%%%%%%%%%%%%%%%%

We now proceed to present numerical results. From now on, we will consider two scenarios, namely, $\{m_{Z'}=m_Q=m_D=m_S=m_T=500\; \mathrm{GeV} \}$ and $\{m_{Z'}=1500\; \mathrm{GeV}, m_Q=m_D=m_S=m_T=700\; \mathrm{GeV} \}$. The results are shown in Table \ref{TABLE3}, where it can be appreciated the more important role played by the exotic quarks. Although the contribution of the third family of known quarks to the $Z'\to ggg$ is dominant, as it occurs for the $Z\to ggg$  transition, it should be noticed that in this case there is a significant contribution from the exotic quarks, which tends to be dominant for a heavier $Z'$ boson. This situation is illustrated in Tables \ref{TABLE4} and \ref{TABLE5} where the contributions arising from the three families as well as the interference effects are shown. In these Tables we also present the values for $\alpha_s$ obtained from Ref.~\cite{PDG}.  On the other hand, the contribution coming from the  exotic quarks is shown with some detail in Table \ref{TABLE6}, in which the interference effects among exotic quarks is shown too. It is important to notice that the individual contribution of the exotic quarks is so important as those of the known quarks, however the global contribution is reduced considerably due to an interference effect between the $D$ and $S$ quarks with the $T$ quark, which is direct consequence of the way in which they appear in the $SU_L(3)$ fundamental representation.

Using for the total decay width of the $Z'$ boson the results given in Ref.~\cite{T1}, the corresponding branching ratio is given by
\begin{equation}
Br(Z'\to ggg)=2.15\times 10^{-5}
\end{equation}
for the scenario characterized by a mass of $m_{Z'}=500$ GeV and
\begin{equation}
Br(Z'\to ggg)=4.95\times 10^{-5}
\end{equation}
for the scenario with $m_{Z'}=1500$ GeV.

\begin{table}
\caption{\label{TABLE3} Partial and total decay widths for the scenarios $\{m_{Z'}=m_Q=m_D=m_S=m_T=500\; \mathrm{GeV} \}$ and
$\{m_{Z'}=1500\; \mathrm{GeV}, m_Q=m_D=m_S=m_T=700\; \mathrm{GeV} \}$.}
 \begin{ruledtabular}
 \begin{tabular}{|l|l|l|l|l|l|l|}
  $m_{Z'}$ [GeV] & $m_Q$ [GeV] & $\alpha_s$ & $\Gamma_{q_i}$ [GeV] & $\Gamma_{Q_i}$ [GeV] & $\Gamma_{q_i-Q_i}$ [GeV] & $\Gamma(Z'\to ggg)$ [GeV]\\
  \hline
  $500$ & $500$ & $0.104482$& $2.74\times 10^{-3}$ & $1.33\times 10^{-8}$& $8.97\times 10^{-6}$& $2.73\times 10^{-3}$ \\
 \hline
  $1500$ &$700$ &$0.150079$ & $2.27\times 10^{-2}$ & $2.8\times 10^{-3}$& $9.11\times 10^{-3}$ & $3.46\times 10^{-2}$
\end{tabular}
\end{ruledtabular}
\end{table}

\begin{table}
 \caption{\label{TABLE4} Family contribution to the $\Gamma(Z'\to ggg)$ decay in the scenario $m_{Z'}=500\; \mathrm{GeV}$ . Here, $\Gamma^{VI}$ and $\Gamma^{AI}$ represent the interference effect induced by the three families into the vector and axial vector width decays, respectively.}
\begin{ruledtabular}
 \begin{tabular}{|l|l|l|l|l|l|}
  Family & $\Gamma^V$ [GeV] & $\Gamma^A$ [GeV] & $\Gamma^{VI}$ [GeV] & $\Gamma^{AI}$ [GeV] & $\Gamma_{q_i}$ [GeV] \\
  \hline
  $u,d$ & $2.24\times 10^{-4}$ & $4.19\times 10^{-4}$ & - & -&- \\
  \hline
  $c,s$ & $2.22\times 10^{-4}$ & $4.36\times 10^{-4}$ & -&- &- \\
  \hline
  $t,b$ & $7.18\times 10^{-4}$ & $4.05\times 10^{-4}$ &- & - & -\\
  \hline
  Total & $1.16\times 10^{-3}$ & $1.26\times 10^{-3}$ & $1.05\times 10^{-3}$ & $-7.39\times 10^{-4}$ & $2.73\times 10^{-3}$
\end{tabular}
\end{ruledtabular}
\end{table}

\begin{table}
 \caption{\label{TABLE5} Family contribution to the $\Gamma(Z'\to ggg)$ decay in the scenario $m_{Z'}=1500\; \mathrm{GeV}$. Here, $\Gamma^{VI}$ and $\Gamma^{AI}$ represent the interference effect induced by the three families into the vector and axial vector width decays, respectively.}
\begin{ruledtabular}
 \begin{tabular}{|l|l|l|l|l|l|}
  Family & $\Gamma^V$ [GeV] & $\Gamma^A$ [GeV] & $\Gamma^{VI}$ [GeV] & $\Gamma^{AI}$ [GeV] & $\Gamma_{q_i}$ [GeV] \\
  \hline
  $u,d$ & $2\times 10^{-3}$    & $3.73\times 10^{-3}$ & - & -&- \\
 \hline
  $c,s$ & $1.98\times 10^{-3}$ & $3.78\times 10^{-3}$ & -&- &- \\
  \hline
  $t,b$ & $2.34\times 10^{-2}$ & $3.44\times 10^{-3}$ &- & - & -\\
  \hline
  Total & $2.74\times 10^{-2}$ & $1.09\times 10^{-2}$ & $-8.85\times 10^{-3}$ & $-6.74\times 10^{-3}$ & $2.27\times 10^{-2}$
\end{tabular}
\end{ruledtabular}
\end{table}

\begin{table}
\caption{\label{TABLE6} Exotic quark contribution to the $Z'\to ggg$ decay in the scenario $\{m_{Z'}=1500\; GeV, m_Q=m_D=m_S=m_T=700\; \mathrm{GeV} \}$.}
\begin{ruledtabular}
\begin{center}
 \begin{tabular}{|l|c|c|c|c|c|c|c|c|}
  Quark & $\Gamma^V_{Q_i}$ [GeV] & $\Gamma^V_{Q_{D-S}}$ [GeV] & $\Gamma^V_{Q_{D-T}}$ [GeV] & $\Gamma^V_{Q_{S-T}}$ [GeV] & $\Gamma^A_{Q_i}$ [GeV] & $\Gamma^A_{Q_{D-S}}$ [GeV] & $\Gamma^A_{Q_{D-T}}$ [GeV] & $\Gamma^A_{Q_{S-T}}$ [GeV] \\
  \hline
  $D$   & $8.58\times 10^{-3}$ & - & - & - & $6.05\times 10^{-6}$ & - & - & -\\
  \hline
  $S$   & $8.58\times 10^{-3}$ & - & - & - & $6.05\times 10^{-6}$ & - & - & -\\
  \hline
  $T$   & $1.76\times 10^{-2}$ & - & - & - & $6.05\times 10^{-6}$ & - & - & -\\
  \hline
  $D,S$ & - &  $1.71\times10^{-2}$  & - & - & - & $1.21\times10^{-5}$ & - & -\\
  \hline
  $D,T$ & - &  -  & $-2.45\times 10^{-2}$ & - & - & - & $-1.21\times10^{-5}$ & -\\
  \hline
  $S,T$ & - & - & - & $-2.45\times 10^{-2}$ & - & - & - & $-1.21\times10^{-5}$\\
\end{tabular}
\end{center}
\end{ruledtabular}
\end{table}

\section{Summary}
\label{co}In this paper, a comprehensive analysis of the rare $Z\to ggg$ and $Z'\to ggg$ decays in the context of the minimal $331$ model has been presented. Explicit expressions for the amplitudes generated at the one-loop level given in terms of Passarino-Veltman scalar functions are presented. The fact that the $Vggg$ vertex ($V=Z,Z'$) is governed by the Bose symmetry is exploited to write its associated vertex function in a compact and manifest $SU_C(3)$-invariant way. The total amplitude is composed by the vector amplitude and the axial vector amplitude, which are finite and gauge-invariant by themselves and do not interfere among themselves, as they are proportional to the color structures $d_{abc}$ and $f_{abc}$, respectively. While the axial vector amplitude receives contributions from both box and triangle diagrams and can be expressed in terms of 24 form factors, the vector amplitude arises only from box diagrams and comprises 18 form factors. It turns out to be that each type of diagrams (box or triangle) leads to amplitudes which are free of ultraviolet divergences and satisfy Bose symmetry. However, in the case of the axial vector amplitude, gauge invariance is obtained only after summing over the contributions arising from box and triangle diagrams. It is found that the vector amplitude also satisfies the transversality conditions with respect to the $V$ vector boson, which means that in this amplitude this vector boson appears only through of the $V_{\mu \nu}=\partial_\mu V_\nu-\partial_\nu V_\mu$ tensor field. This property is not present in the axial vector amplitude, which is transverse only with respect to the gluonic fields. Our results are valid for any renormalizable theory and are model-independent in this sense.

As far as the numerical results is concerned, the behavior of the vector and axial vector amplitudes are analyzed as a function of the mass quark and also as a function of the mass difference of the members of the quark family. It was found that both type of amplitudes show a decoupling nature with respect to the former case, whereas a nondecoupling behavior is shown with respect to the latter case. In the case of the $Z\to ggg$ decay, the axial vector amplitude vanishes in the degenerate case and reach its maximum value for the third family. The axial vector contribution to this decay is marginal, as it is almost one order of magnitude lower than that associated to the vector amplitude. This decay is insensitive to the presence of exotic quarks, as it is essentially governed by the third family, especially by the bottom quark, whose branching ratio is given by $Br(Z\to ggg)=1.4\times 10^{-5}$. All the results given in the literature were nicely reproduced. As to the $Z'\to ggg$ decay, its behavior present some differences with respect to the standard $Z\to ggg$ decay, as it couples differently to the SM quarks. In particular, its couplings to the third family of quarks differs of its couplings to the first and second families, as in the $331$ model the former is accommodate as an antitriplet of $SU_L(3)$, whereas the latter two are introduced as triplets of this group. In this case, the axial vector amplitude does not vanish in the degenerate case and its contribution is, in some scenarios, as important as the one given by the vector amplitude. In a scenario with $m_{Z'}=500$ GeV, the three families give contributions of the same order of magnitude to both the vector amplitude and the axial vector amplitude. The situation changes substantially for a heavier $Z'$ boson, as the vector amplitude receives a dominant contribution from the third family, especially from the top quark. In this case, the contribution of the exotic quarks is much less marginal than in the case of the $Z\to ggg$ decay, and tends to assume a dominant role for a heavier $Z'$ boson. Although the separate contribution of each exotic quark is so important as the one arising from the known quarks, there is an interference effect between the $D$ and $S$ quarks with the $T$ quark that reduce their global contribution by about one order of magnitude. For instance, in a scenario with $m_{Z'}=1500$ GeV, this contribution is one order of magnitude lower than that arising from the known quarks, but it tends to increases with the $Z'$ mass. In this scenario, the contribution of the third family to the vector amplitude is one order of magnitude larger than the corresponding contribution of the other two families and also one order of magnitude larger than the axial vector component of the decay width, which receives contributions of the same order of magnitude from the three families. Thus, while the $Z\to ggg$ decay is governed by the third family, the $Z'\to ggg$ one receives important contributions from the three families. The contribution of exotic quarks to the $Z\to ggg$ decay is completely marginal, but they play an significant role in the case of the  $Z'\to ggg$ decay, especially for a relatively heavy $Z'$ boson. In general terms, the decay width for $Z'\to ggg$ is almost three orders of magnitude larger than that for $Z\to ggg$. Also, the $Z'\to ggg$ decay has a branching ratio larger than the $Z\to ggg$ decay, which is of $Br(Z'\to ggg)=2.15\times 10^{-5}$ and $Br(Z'\to ggg)=4.95\times 10^{-5}$ for $m_{Z'}=500$ GeV and $m_{Z'}=1500$ GeV, respectively.

\acknowledgments{We acknowledge financial support from CONACYT and
SNI (M\' exico).}

\appendix*

\section{Form factors of the $Vggg$ vertex}

The 3 representative vector form factors are given by

\begin{eqnarray}
f_{V1}^{q}&=&-\frac{B_0(1) (p_{13}-2 p_{23})}{6 p_{13}^2 p_{23}}+\frac{B_0(3) (2 p_{12}-p_{13}) p_{23}}{6
p_{12}^2 p_{13}^2}-\frac{B_0(2) (p_{12}+p_{23})}{6 p_{12}^2 p_{23}}-\frac{C_0(2) p_{13} (2 p_{12}^3+3 p_{23}^2 p_{12}+2 p_{23}^3)}{12 p_{12}^3 p_{23}^2}\nonumber\\&&
+\frac{B_0(4) (p_{12}+p_{13}+p_{23}) [p_{12} (p_{13}-2 p_{23})+p_{13} p_{23}]}{6 p_{12}^2 p_{13}^2 p_{23}}
+\frac{D_0(1) [2 p_{23}^2 m_{q}^4+p_{12} (2 p_{13}-3 p_{23}) p_{23} m_{q}^2+2 p_{12}^2 p_{13}^2]}{6
p_{12} p_{13} p_{23}^2}\nonumber\\ &&
+\frac{C_0(4) (p_{13}+p_{23}) (2 p_{13}^3-3 p_{23}^2 p_{13}-4 p_{23}^3)}{12 p_{13}^3 p_{23}^2}
+\frac{C_0(5) (p_{12}+p_{23}) [2 p_{13} p_{12}^3+3 (2 m_{q}^2+p_{13}) p_{23}^2 p_{12}+2 p_{13} p_{23}^3]}{12
p_{12}^3 p_{13} p_{23}^2} \nonumber\\ &&
-\frac{C_0(6) (p_{12}+p_{13}) [(3 p_{13}+4 p_{23}) p_{12}^3-3 p_{13}^3 p_{12}-2 p_{13}^3 p_{23}]}{12 p_{12}^3 p_{13}^3}
+C_0(1) \left[\frac{1}{12} p_{12} \left(\frac{3 p_{13}+4 p_{23}}{p_{13}^3}-\frac{2}{p_{23}^2}\right)-\frac{m_{q}^2}{2 p_{12} p_{13}}\right]\nonumber\\&&
+\frac{C_0(3) p_{23} [(3 p_{13}+4 p_{23}) p_{12}^3-3 p_{13}^2 (2 m_{q}^2+p_{13}) p_{12}-2 p_{13}^3
p_{23}]}{12 p_{12}^3 p_{13}^3}-\frac{1}{6 p_{12} p_{13}}\nonumber\\ &&
+\frac{D_0(3) [2 p_{12}^2 m_{q}^4+p_{12} (-3 p_{12}^2+3 p_{13} p_{12}+5
p_{13} p_{23}) m_{q}^2+p_{13}^2 p_{23} (3 p_{12}+2 p_{23})]}{6 p_{12}^3 p_{13}}\nonumber\\&&
+\frac{D_0(2) \{2 p_{13}^2 m_{q}^4+p_{13} [3 p_{13} (p_{13}+p_{23})-p_{12} (3 p_{13}+4 p_{23})] m_{q}^2-p_{12}^2
p_{23} (3 p_{13}+4 p_{23})\}}{6 p_{12} p_{13}^3}\;,
\end{eqnarray}

\begin{eqnarray}
f_{V7}^{q}&=&\frac{D_0(2)  m_{q}^2}{2 p_{23}^2}\left(\frac{2 m_{q}^2}{p_{12}}+\frac{p_{23}}{p_{13}}\right)+
\frac{D_0(3) [2p_{12} m_{q}^2+(p_{12}+2 p_{13}) p_{23}] m_{q}^2}{2 p_{12}^2 p_{23}^2}
-\frac{C_0(3) m_{q}^2}{2 p_{12}^2p_{23}}+\frac{C_0(6) (p_{12}+p_{13})^2 m_{q}^2}{2 p_{12}^2 p_{13} p_{23}^2}\nonumber\\&&
+ \frac{B_0(4) (p_{12}+p_{13}+p_{23}) (2 p_{12}+3 p_{23})}{2 p_{23}^3 (p_{12}+p_{23})^2}-\frac{C_0(1)
(2 p_{13}+p_{23}) (p_{23} m_{q}^2+p_{12} p_{13})}{2 p_{13} p_{23}^4}+\frac{1}{2 p_{23}^2 (p_{12}+p_{23})}-\frac{B_0(1)}{p_{23}^3}\nonumber\\&&
+\frac{C_0(4) (p_{13}+p_{23}) [2 p_{23} m_{q}^2+p_{12} (2 p_{13}+p_{23})]}{2 p_{12} p_{23}^4}
-\frac{B_0(2) [p_{12} (2 p_{13}+p_{23})+p_{23} (3 p_{13}+p_{23})]}{2 p_{23}^3 (p_{12}+p_{23})^2}\nonumber\\&&
+\frac{C_0(5) [(2 p_{13}+p_{23}) p_{12}^4+2 p_{23} (m_{q}^2+2 p_{13}+p_{23}) p_{12}^3+p_{23}^2 (3 m_{q}^2+2 p_{13}+p_{23}) p_{12}^2+m_{q}^2 p_{23}^4]}{2 p_{12}^2 p_{23}^4 (p_{12}+p_{23})}\nonumber\\&&
+\frac{D_0(1)[2 p_{23}^2 m_{q}^4+p_{12} p_{23} (8 p_{13}+3 p_{23}) m_{q}^2+2 p_{12}^2 p_{13} (2 p_{13}+p_{23})]}{2p_{12} p_{23}^4}\nonumber\\&&
-\frac{C_0(2) [p_{13} p_{23}^2 m_{q}^2+p_{12} p_{23} (2 p_{13}+p_{23}) m_{q}^2+p_{12}^2 p_{13}
(2 p_{13}+p_{23})]}{2 p_{12}^2 p_{23}^4} \;,
\end{eqnarray}

\begin{eqnarray}
f_{V13}^{q}&=&\frac{B_0(3) (4 p_{12}+p_{13})}{12 p_{12} p_{13}^2}-\frac{C_0(6) (p_{12}+p_{13}) (4 p_{12}^3+p_{13}^3)}{12 p_{12}^2 p_{13}^3}+\frac{B_0(2) (p_{23}-8 p_{12})}{12 p_{12} p_{23}^2}+C_0(3) \left(\frac{p_{23}}{12 p_{12}^2}+\frac{p_{12} p_{23}}{3 p_{13}^3}\right)\nonumber\\&&
+ \frac{D_0(2) (p_{13}^2 m_{q}^4-2 p_{12} p_{13} p_{23} m_{q}^2-2 p_{12}^2 p_{23}^2)}{3 p_{13}^3 p_{23}}
+\frac{D_0(3) [2 p_{12}^2 m_{q}^4+p_{12} (3 p_{12}-p_{13}) p_{23} m_{q}^2-p_{13}^2 p_{23}^2]}{6 p_{12}^2 p_{13} p_{23}}\nonumber\\&&
-\frac{C_0(1) p_{12} [8 p_{12} p_{13}^3+3 (2 m_{q}^2+p_{13}) p_{23} p_{13}^2-4 p_{12} p_{23}^3]}{12
p_{13}^3 p_{23}^3}+\frac{2 p_{13}-p_{23}}{6
p_{23} p_{13}^2+6 p_{23}^2 p_{13}}\nonumber\\&&+
\frac{C_0(2) [-8 p_{13} p_{12}^3-3 (2 m_{q}^2+p_{13}) p_{23} p_{12}^2+p_{13} p_{23}^3]}{12 p_{12}^2
p_{23}^3}+\frac{C_0(5) (p_{12}+p_{23}) [8 p_{13} p_{12}^3+3 (2 m_{q}^2+p_{13}) p_{23} p_{12}^2-p_{13}
p_{23}^3]}{12 p_{12}^2 p_{13} p_{23}^3}\nonumber\\&&
+\frac{B_0(1) [-8 p_{12} p_{13}^3-3 (4 p_{12}+p_{13}) p_{23} p_{13}^2+6 p_{12} p_{23}^2
p_{13}+(4 p_{12}+3 p_{13}) p_{23}^3]}{12 p_{13}^2 p_{23}^2 (p_{13}+p_{23})^2}\nonumber\\&&
+\frac{B_0(4) (p_{12}+p_{13}+p_{23}) [2 p_{12} (4 p_{13}^3+6 p_{23} p_{13}^2-3 p_{23}^2 p_{13}-2 p_{23}^3)-p_{13}
p_{23} (p_{13}+p_{23})^2]}{12 p_{12} p_{13}^2 p_{23}^2 (p_{13}+p_{23})^2}\nonumber\\&&
+\frac{D_0(1) \{8 p_{12}^2 p_{13}^2+p_{12} (14 m_{q}^2+3 p_{13}) p_{23} p_{13}+m_{q}^2 p_{23}^2 [2 m_{q}^2+3 (p_{13}+p_{23})]\}}{6 p_{13} p_{23}^3}\nonumber\\&&
+\frac{C_0(4) \{3 p_{23} [2 (p_{13}^2+2 p_{23} p_{13}-p_{23}^2) m_{q}^2+p_{13} (p_{13}+p_{23})^2] p_{13}^2+4
p_{12} (p_{13}+p_{23})^2 (2 p_{13}^3-p_{23}^3)\}}{12 p_{13}^3 p_{23}^3 (p_{13}+p_{23})} \;.
\end{eqnarray}

The 4 representative axial vector form factors are given by

\begin{eqnarray}
f_{A 1}^{q} & = & \frac{1}{4 p_{12}^3} \bigg\{ - \frac{2 p_{12}^2}{p_{12} + p_{13}} + 2 p_{12} [B_0(2) - B_0(4)] + \frac{2 p_{12} [p_{13} p_{23} + p_{12} (p_{12} + p_{13} + 2 p_{23})] [B_0(3) - B_0(4)]}{(p_{12} + p_{13})^2} \nonumber \\
& & + (p_{12} + 2 p_{23}) [p_{13} C_0(2) + p_{23} C_0(3)] - \frac{(p_{12} + p_{23}) [2 m_{q}^2 p_{12} + p_{13} (p_{12} + 2 p_{23})] C_0(5)}{p_{13}} \nonumber \\ & & + \frac{2 m_{q}^2 p_{12} (p_{13} + p_{23}) C_0(4)}{p_{13}} - \frac{[2 m_{q}^2 p_{12} (p_{13}^2 + 2 p_{12} p_{13} - p_{12}^2) + p_{13} (p_{12} + p_{13})^2 (p_{12} + 2 p_{23})] C_0(6)}{p_{13} (p_{12} + p_{13})}\nonumber  \\
& & + \frac{2 m_{q}^2 p_{12}^2 (p_{13} + p_{23}) D_0(2)}{p_{13}} - 2 [m_{q}^2 p_{12} (p_{12} + 3 p_{23}) + p_{13} p_{23} (p_{12} + 2 p_{23})] D_0(3) \bigg\} \,
\end{eqnarray}

\begin{eqnarray}
f_{A 7}^{q} & = & \frac{1}{2 p_{12} p_{13}^3} \bigg\{ \frac{p_{13}^2 (p_{13}^2 - p_{12} p_{23})}{p_{23} (p_{12} + p_{13}) (p_{13} + p_{23})} + \frac{p_{12} p_{13} (2 p_{13} + p_{23}) [B_0(1) - B_0(4)]}{(p_{13} + p_{23})^2} + \frac{p_{12} p_{13} (p_{12} + 2 p_{13}) [B_0(3) - B_0(4)]}{(p_{12} + p_{13})^2}\nonumber \\
& & + p_{12}^2 C_0(1) + p_{12} p_{23} C_0(3) - \frac{[m_{q}^2 p_{13} (p_{23}^2 + 2 p_{13} p_{23} - p_{13}^2) + p_{12} p_{23} (p_{13} + p_{23})^2] C_0(4)}{p_{23} (p_{13} + p_{23})}\nonumber \\
& & + \frac{m_{q}^2 p_{13} (p_{12} + p_{23}) C_0(5)}{p_{23}} - \frac{[m_{q}^2 p_{13} (p_{12}^2 + 2 p_{12} p_{13} - p_{13}^2) + p_{12} p_{23} (p_{12} + p_{13})^2] C_0(6)}{p_{23} (p_{12} + p_{13})} + \frac{m_{q}^2 p_{12} p_{13}^2 D_0(1)}{p_{23}}\nonumber \\
& & - p_{12} (3 m_{q}^2 p_{13} + 2 p_{12} p_{23}) D_0(2) + m_{q}^2 p_{13}^2 D_0(3) \bigg\} \ ,
\end{eqnarray}

\begin{eqnarray}
f_{A 13}^{q} & = & \frac{1}{4} \bigg\{ \frac{2 [p_{13} p_{23} + p_{12} (p_{13} + 2 p_{23})]}{p_{12} p_{23} (p_{12} + p_{13}) (p_{13} + p_{23})} + \frac{(p_{13}^2 - 2 p_{13} p_{23} - p_{23}^2) [B_0(1) - B_0(4)]}{p_{13} p_{23} (p_{13} + p_{23})^2} - \frac{B_0(2) - B_0(4)}{p_{12} p_{23}}\nonumber \\
& & - \frac{(p_{12}^2 + 4 p_{12} p_{13} + p_{13}^2) [B_0(3) - B_0(4)]}{p_{12} p_{13} (p_{12} + p_{13})^2} - \frac{p_{12} C_0(1)}{p_{13}^2} - \frac{p_{13} C_0(2)}{p_{12}^2} - \frac{p_{23} (p_{12}^2 + p_{13}^2) C_0(3)}{p_{12}^2 p_{13}^2}\nonumber \\
& & + \frac{[4 m_{q}^2 p_{13}^2 + p_{12} (p_{13} + p_{23})^2] C_0(4)}{p_{12} p_{13}^2 (p_{13} + p_{23})} + \frac{(p_{12} + p_{23}) C_0(5)}{p_{12}^2} + \frac{[4 m_{q}^2 p_{12}^2 p_{13}^2 + p_{23} (p_{12} + p_{13})^2 (p_{12}^2 + p_{13}^2)] C_0(6)}{p_{23} p_{12}^2 p_{13}^2 (p_{12} + p_{13})} \nonumber \\
& & + \frac{2 [m_{q}^2 p_{13} (p_{13} + p_{23}) + p_{12} p_{23}^2] D_0(2)}{p_{23} p_{13}^2} - \frac{2 [m_{q}^2 p_{12} (p_{12} - p_{23}) - p_{13} p_{23}^2] D_0(3)}{p_{23} p_{12}^2} \bigg\} \ ,
\end{eqnarray}

\begin{eqnarray}
f_{A 19}^{q} & = & \frac{1}{4} \bigg\{ - \frac{2}{p_{23} (p_{12} + p_{13})} + \frac{2 [B_0(2) - B_0(4)]}{p_{12} (p_{12} + p_{23})} + \frac{2 [B_0(3) - B_0(4)]}{(p_{12} + p_{13})^2} + \frac{p_{13} C_0(2)}{p_{12}^2} + \frac{p_{23} C_0(3)}{p_{12}^2} - \frac{(p_{12} + p_{23}) C_0(5)}{p_{12}^2}\nonumber \\
& & - \frac{[4 m_{q}^2 p_{12}^2 + p_{23} (p_{12} + p_{13})^2] C_0(6)}{p_{23} p_{12}^2 (p_{12} + p_{13})} + \frac{2 m_{q}^2 [D_0(1) - D_0(2)]}{p_{12}} - \frac{2 (m_{q}^2 p_{12} + p_{13} p_{23}) D_0(3)}{p_{12}^2} \bigg\} \ .
\end{eqnarray}

In writing the above expressions we have introduced the following definitions:

\begin{eqnarray}
B_0(1)&\equiv& B_0(2 p_{12},m_{q}^2,m_{q}^2), \nonumber\\
B_0(2)&\equiv& B_0(2 p_{13},m_{q}^2,m_{q}^2), \nonumber\\
B_0(3)&\equiv& B_0(2 p_{23},m_{q}^2,m_{q}^2), \nonumber\\
B_0(4)&\equiv& B_0(m_V^2,m_{q}^2,m_{q}^2), \nonumber
\end{eqnarray}
\begin{eqnarray}
C_0(1)&\equiv& C_0(0,0,2 p_{12},m_{q}^2,m_{q}^2,m_{q}^2), \nonumber\\
C_0(2)&\equiv& C_0(0,0,2 p_{13},m_{q}^2,m_{q}^2,m_{q}^2), \nonumber\\
C_0(3)&\equiv& C_0(0,0,2 p_{23},m_{q}^2,m_{q}^2,m_{q}^2), \nonumber\\
C_0(4)&\equiv& C_0(0,2 p_{12},m_V^2,m_{q}^2,m_{q}^2,m_{q}^2), \nonumber\\
C_0(5)&\equiv& C_0(0,2 p_{13},m_V^2,m_{q}^2,m_{q}^2,m_{q}^2), \nonumber\\
C_0(6)&\equiv& C_0(0,2 p_{23},m_V^2,m_{q}^2,m_{q}^2,m_{q}^2), \nonumber
\end{eqnarray}
\begin{eqnarray}
D_0(1)&\equiv& D_0(0,0,0,m_V^2,2 p_{12},2 p_{13},m_{q}^2,m_{q}^2,m_{q}^2,m_{q}^2), \nonumber\\
D_0(2)&\equiv& D_0(0,0,0,m_V^2,2 p_{12},2 p_{23},m_{q}^2,m_{q}^2,m_{q}^2,m_{q}^2), \nonumber\\
D_0(3)&\equiv& D_0(0,0,0,m_V^2,2 p_{13},2 p_{23},m_{q}^2,m_{q}^2,m_{q}^2,m_{q}^2), \nonumber
\end{eqnarray}
where $p_{ij}\equiv p_i\cdot p_j$, with $i,j$=1,2,3.

\end{document}